\documentclass{article}

\usepackage[numbers,square,sort]{natbib}
\usepackage[preprint]{neurips_2022}
\usepackage[numbers,square,sort]{natbib}

\usepackage{etoolbox}
\newtoggle{nmi}
\togglefalse{nmi}

\usepackage{graphicx}
\usepackage{float}
\usepackage{lineno}
\usepackage{xcolor}
\usepackage{float} %

\newcommand{\jr}[2]{\textcolor{black}{#1}}
\newcommand{\jrr}[2]{\textcolor{black}{#1}}
\newcommand{\red}{\color{black}}
\newcommand{\redd}{\color{black}}

\usepackage{authblk}
\makeatletter
\renewcommand\AB@affilsepx{, \protect\Affilfont}
\makeatother

\usepackage[utf8]{inputenc} %
\usepackage[T1]{fontenc}    %
\usepackage{hyperref}       %
\usepackage{url}            %
\usepackage{booktabs}       %
\usepackage{amsfonts}       %
\usepackage{nicefrac}       %
\usepackage{microtype}      %
\usepackage{xcolor}         %
\usepackage{graphicx}
\usepackage{scalerel}
\usepackage{caption}
\usepackage{subcaption}

\usepackage{pgfplots}
\usepackage{pgf}
\usepackage{tikzscale}
\usepackage{tikz}
\pgfplotsset{compat=newest}
\usetikzlibrary{arrows,calc,positioning,shapes.geometric,math}
\makeatletter
\newcommand\tikzcurr{\the\tikz@lastxsaved,\the\tikz@lastysaved}
\makeatother

\usepackage{etoolbox}
\newtoggle{release}
\togglefalse{release}
\iftoggle{release}{
\newcommand{\alex}[1]{}
\newcommand{\jremi}[1]{}
\newcommand{\charl}[1]{}
}
{
\newcommand{\alex}[1]{{\color{blue} A: #1}}
\newcommand{\jremi}[1]{{\color{green} J: #1}}
\newcommand{\charl}[1]{{\color{magenta}C: #1}}
}

\usepackage{amsmath, amssymb, amsthm}
\usepackage{bm,bbm}
\newcommand{\proba}[1]{\mathbb{P}\left[#1\right]}
\newcommand{\dotprod}[2]{\langle #1{,}#2\rangle}
\newcommand{\reel}{\mathbb{R}}

\newcommand{\complex}{\mathbb{C}}

\newcommand{\decreg}{{\bm{f}_\mathrm{reg}}}
\newcommand{\enc}{{\bm{f}_\mathrm{clip}}}

\newcommand{\ee}{\mathrm{e}}

\title{Decoding speech \jr{perception}{} from non-invasive brain recordings}

\author[1,*]{Alexandre Défossez}
\author[1,2]{Charlotte Caucheteux}
\author[1]{Jérémy Rapin}
\author[1]{Ori Kabeli}
\author[1,3,*]{Jean-Rémi King}

\affil[1]{Meta AI}
\affil[2]{Inria Saclay}
\affil[2]{PSL University}
\affil[*]{\texttt{\{defossez;jeanremi\}@meta.com}}

\begin{document}

\maketitle

\begin{abstract}
\looseness=-1
Decoding speech from brain activity is a long-awaited goal in both healthcare and neuroscience. Invasive devices have recently led to major milestones in that regard: deep learning algorithms trained on intracranial recordings now start to decode elementary linguistic features (e.g. letters, words, spectrograms). However, extending this approach to natural speech and non-invasive brain recordings remains a major challenge. \jr{Here,}{} we introduce a model trained \jr{with contrastive-learning}{} to decode self-supervised representations of \jr{perceived}{} speech from the non-invasive recordings of a large cohort of \jr{healthy}{} individuals. To evaluate this approach, we curate and integrate four public datasets, encompassing \jr{175}{} volunteers recorded with magneto- or electro-encephalography (M/EEG), while they listened to \jr{short stories and isolated sentences}{}. The results show that our model can identify, from 3 seconds of MEG signals, the corresponding speech segment with up to \jr{41\%}{} accuracy out of \jr{more than 1,000}{} distinct possibilities \jr{on average across participants, and more than 80\% in the very best participants}{} -- a performance that allows the decoding of \jrr{words and}{} phrases absent from the training set. \jr{The comparison of our model to a variety of baselines highlights the importance}{} of (i) a contrastive objective, (ii) pretrained representations of speech and (iii) a common convolutional architecture simultaneously trained across multiple participants. \jr{Finally, the analysis of the decoder’s predictions suggests that they primarily depend on lexical and contextual semantic representations.}{} Overall, \jr{this effective decoding of perceived speech from non-invasive recordings delineates}{} a promising path \jr{to decode language from brain activity, without putting patients}{} at risk for brain surgery.

\end{abstract}

\section{Introduction}

Every year\jr{, traumatic brain injuries, strokes and neurodegenerative diseases make thousands of patients}{} lose their ability to \jr{speak or even communicate}{}~\citep{stanger1996demographics,pels2017estimated,kubler2001brain,pels2017estimated,claassen2019detection,owen2006detecting,cruse2011bedside}. 
Brain Computer Interface (BCI) has been raising high expectations to detect \citep{owen2006detecting,claassen2019detection,birbaumer1999spelling,king2013single} and restore \jr{communication}{language} abilities in such patients \citep{brumberg2009artificial,herff2015brain,stavisky2018decoding,willett2021high,moses2021neuroprosthesis,kennedy2022slow}.  
Over the past decades, several teams used BCI to efficiently decode phonemes, speech sounds \citep{pei2011decoding,akbari2019towards}, hand gestures \citep{stavisky2018decoding,willett2021high} and articulatory movements \citep{anumanchipalli2019speech,moses2021neuroprosthesis} from electrodes implanted in the cortex or over its surface. 
For instance, \citet{willett2021high} decoded 90 characters per minute (with a 94\% accuracy, \emph{i.e.} roughly $\approx$15-18 words per minute) from a spinal-cord injury patient recorded in the motor cortex during 10 hours of writing sessions. Similarly, \citet{moses2021neuroprosthesis} decoded 15.2 words per minute (with 74.4\% accuracy, and using a vocabulary of 50 words) in an anarthria patient implanted in the sensorimotor cortex and recorded over 48 sessions spanning over 22 hours.
\jr{Finally, \citet{metzger2022generalizable} recently showed that a patient with severe limb and vocal-tract paralysis and implanted over the sensori-motor cortex could efficiently spell words using a code word that represented each English letter (e.g. 
"alpha" for "a"): this approach leads to 6.13\% character error rate and 29.4\% characters per minute, and hence starts to provide a viable communication channel for such patients.}{}

However, such invasive recordings face a major practical challenge: these high-quality signals require \jr{brain surgery and can be difficult to maintain chronically.}{(1) brain surgery (2) long training sessions and (3) can be difficult to maintain chronically.}
Several laboratories have thus focused on decoding language from \textit{non-invasive} recordingss of brain activity like magneto- and electro-encephalography (M/EEG). MEG and EEG are sensitive to macroscopic changes of electric and magnetic signals elicited in the cortex, and can be acquired with a safe and potentially wearable setup \citep{boto2018moving}. However, these two devices produce notoriously noisy signals that vary greatly across sessions and across individuals \citep{hamalainen1993magnetoencephalography,schirrmeister2017deep,king2018encoding}. It is thus common to engineer pipelines that output hand-crafted features, which, in turn, can be learned by a decoder trained on a single participant \citep{panachakel2021decoding,lawhern2018eegnet,lopopolo2020part,chan2011decoding,nguyen2017inferring,murphy2022decoding}. %

In sum, decoding language from brain activity is, to date, either limited to invasive recordings or to impractical tasks. Interestingly, both of these approaches followed a similar method: \emph{i.e.} (1) training a model on a single patient and (2) aiming to decode a limited set of interpretable features (MEL spectrogram, letters, phonemes, small set of words).

\begin{figure}[ht]
  \centering
  \includegraphics[width=0.9\textwidth]{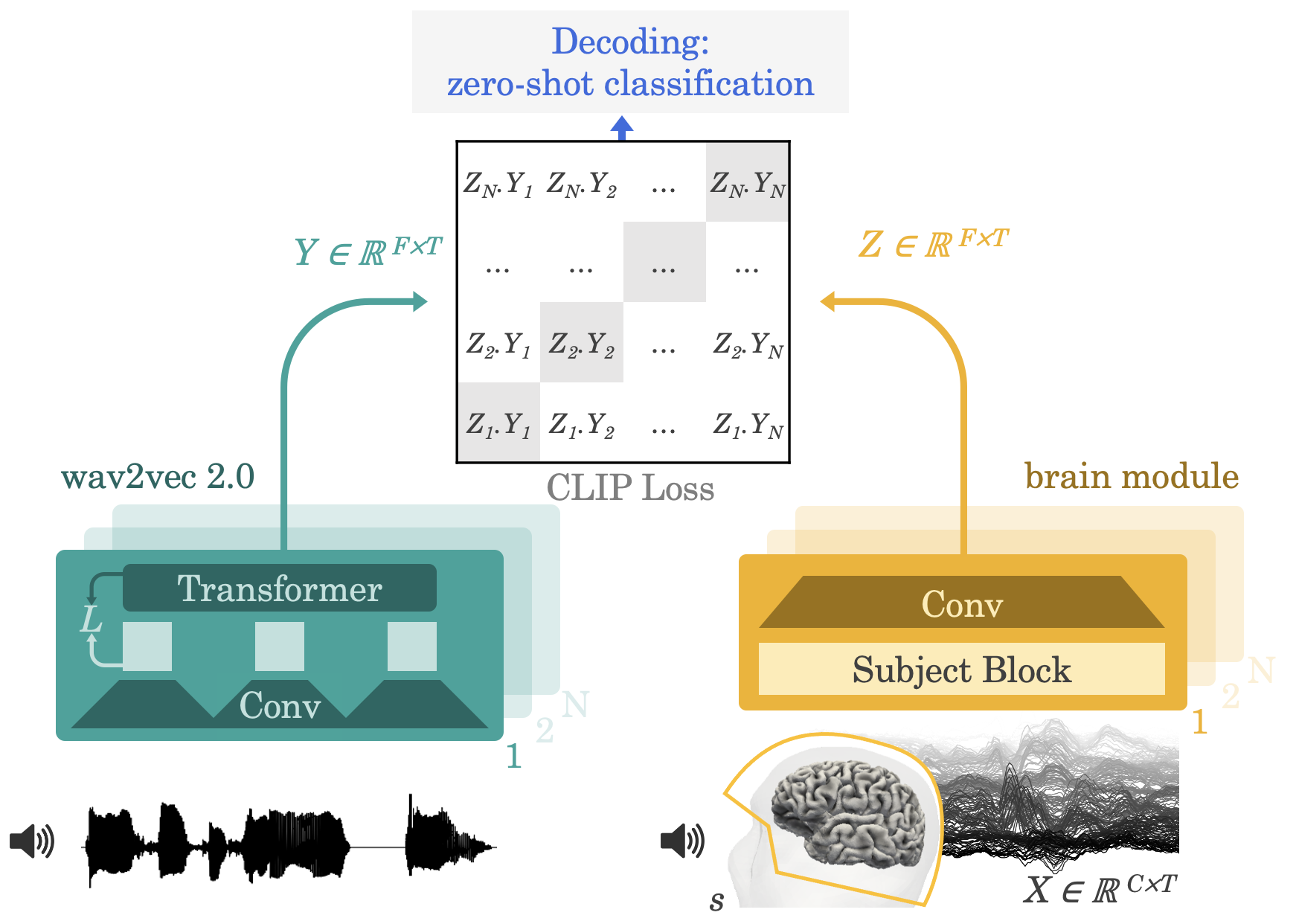}
  \caption{
  \textbf{Method.} 
  We aim to decode speech from the brain activity of healthy participants recorded with magnetoencephalography (MEG) or electroencephalography (EEG) while they listen to stories and/or sentences. For this, our model extracts the deep contextual representations of 3\,s speech signals ($Y$ \jr{of $F$ feature by $T$ time samples}{}) from a pretrained \jr{`speech module`}{self supervised model} (wav2vec 2.0: \citet{baevski2020wav2vec}) and learns the representations ($Z$) of the brain activity on the corresponding 3\,s window ($X$ \jr{of $C$ recording channels by $T$ time samples}{}) 
  that maximally align with these speech representations with a contrastive loss (CLIP: \citet{radford2021learning}). 
  The representation $Z$ is given by a deep convolutional network.
    At evaluation, we input the model with left-out sentences and compute the probability of each 3\,s speech segment given each brain representation. %
    The resulting decoding can thus be ``zero-shot'' in that the audio snippets predicted by the model need not be present in the training set. This approach is thus more general than standard classification approaches where the decoder can only predict the categories learnt during training.
  }
  \label{fig:methods}
\end{figure}

Instead, we here propose to decode speech from non-invasive brain recordings by using 
(1) a single architecture trained across a large cohort of participants and 
(2) deep representations of speech learnt with self-supervised learning on a large quantity of speech data. 
\jr{We focus the present work on speech \emph{perception} in \emph{healthy} volunteers rather than speech \emph{production} in \emph{patients} to design a deep-learning architecture that effectively addresses two core challenges: (1) the fact that non-invasive brain recording can be extremely noisy and variable across trials and participants and (2) the fact that the nature and format of language representations in the brain remain largely unknown.}{}
For this, we introduce a convolutional neural network stacked onto a ``Subject Layer'' and trained with a contrastive objective to predict the representations of the audio waveform learnt by wav2vec 2.0 pretrained on 56k hours of speech \citep{baevski2020wav2vec} (Figure \ref{fig:methods}).
To validate our approach, we curate and integrate four public M/EEG datasets, encompassing the brain activity of \jr{175}{169} participants passively listening to sentences of short stories. 
With a sample of 3\,seconds of M/EEG signals, our model identifies the matching audio segment (\emph{i.e.} zero-shot decoding) with up to 72.5\% top-10 accuracy (out of 1,594 segments) for MEG and up to 19.1\% top-10 accuracy (out of 2,604 segments) for EEG.

\jr{This approach provides three main contributions for the development of non-invasive BCI. First, it shows how pretrained speech models can leverage the decoding of speech in the brain, without exposing volunteers to a tedious repetition of every single word targeted by the decoder. Second, this study shows how our design choices – including contrastive learning and our multi-subject architecture – lead to an efficient processing of continuous EEG and MEG recordings, and thus offers a data-driven guideline for the development of future BCI. Finally, our analyses suggest that the resulting decoder primarily relies on high-level representations of words and phrases.}{}

\section{Method}

\label{sec:method}

We first formalize the general task of neural decoding and then describe and motivate the different components of our model, before describing the datasets, preprocessing, training and evaluation.

\subsection{Problem formalization}
\label{sec:brain_signals}

\begin{figure}[t]
    \centering
    \includegraphics[width=1.\textwidth]{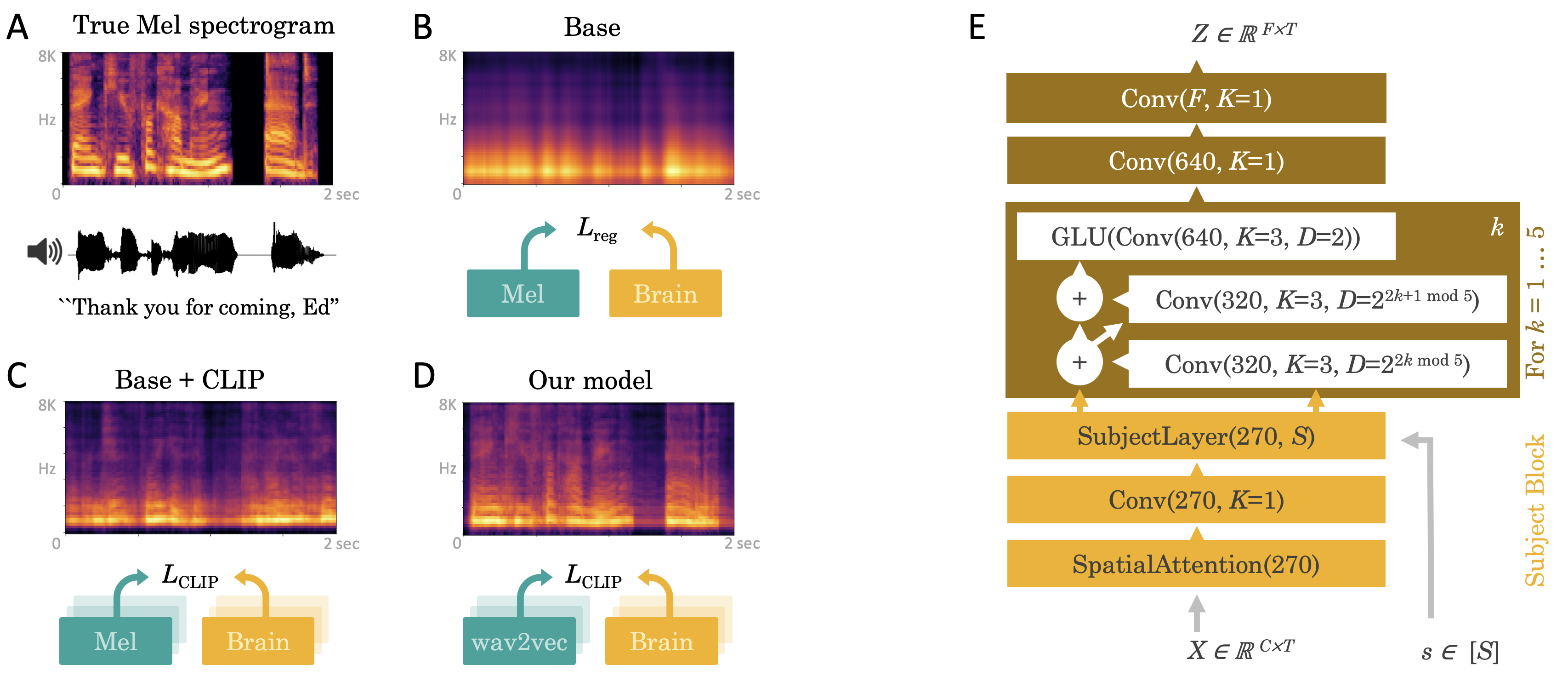}
    \caption{
    \textbf{Design choices. A.} Illustration of a 3\,s speech sound segment (bottom) and its corresponding Mel spectrogram (top).
    \textbf{B.} Mel-spectrogram predicted with a direct regression loss $L_{\mathrm{reg}}$ of a brain decoder (orange).
    \textbf{C.} Replacing the regression loss with a CLIP loss \citep{radford2021learning} improves reconstruction in the same subject,
        still using the mel-spectrogram as the speech representation.
    \textbf{D.} Now replacing the mel-spectrogram with wav2vec 2.0 \citep{baevski2020wav2vec}.
        The probabilities given by Eq. \eqref{eq:clip_prob} are used to rebuild a mel-spectrogram.
    \textbf{E. Architecture of the brain module.} %
    Architecture used to process the brain recordings.
    For each layer, we note first the number of output channels, while the number of time steps is constant throughout the layers. The model is composed of a spatial attention layer, then a 1x1 convolution without activation. 
    A ``Subject Layer'' is selected based on the subject index $s$, which consists in a 1x1 convolution learnt only for that subject with no activation.
    Then, we apply five convolutional blocks made of three convolutions.
    The first two use residual skip connection and increasing dilation, followed by a BatchNorm layer and a GELU activation.
    The third convolution is not residual, and uses a GLU activation (which halves the number of channels) and no normalization.
    Finally, we apply two 1x1 convolutions with a GELU in between.
    }
    \label{fig:old_vs_new}
\end{figure}

We aim to decode speech from a time series of high-dimensional brain signals recorded with non-invasive magneto-encephalography (MEG) or electro-encephalography (EEG) while healthy volunteers passively listened to spoken sentences in their native language. How spoken words are represented in the brain is largely unknown \citep{hickok2007cortical,huth2016natural,caucheteux2022deep}. Thus, it is common to train decoders in a supervised manner to predict a latent representation of speech known to be relevant to the brain \citep{akbari2019towards,angrick2019speech,angrick2019interpretation,krishna2020speech,komeiji2022transformer}. For example, the Mel spectrogram is often targeted for neural decoding because it \jr{represents}{} sounds similarly to the cochlea~\citep{mermelstein_mel}. 
\jr{We formalize this problem as follows.} {}
Let $X \in \reel^{C\times T}$ be a segment of a brain recording of a given subject while \jr{she}{they} listens to a speech segment of the same duration, with $C$ the number of M/EEG sensors and $T$ the number of time steps.
Let $Y \in \reel^{F\times T}$ be the latent representation of speech, using the same sample rate as $X$ for simplicity, here the Mel spectrogram with $F$ frequency bands. \jr{In this formalization}{Thus}, supervised decoding consists of finding a decoding function:
$
\decreg: \reel^{C\times T} \rightarrow \reel^{F\times T}
$
such that $\decreg$ predicts $Y$ given $X$.
We denote by $\hat{Y} = \decreg(X)$ the representation of speech decoded from the brain.
When $\decreg$ belongs to a parameterized family of models like deep neural networks, it can be trained with a regression loss $L_{\mathrm{reg}}(Y, \hat{Y})$ (\emph{e.g.} the Mean Square Error),
\begin{linenomath}
\begin{equation}
\label{eq:regression}
   \min_\decreg \sum_{X, Y} L_{\mathrm{reg}}(Y, \decreg(X)).
\end{equation}
\end{linenomath}
\jr{This}{Empirically, we observed that this} direct regression approach \jr{}{faces several challenges: decoding predictions} appears to be dominated by a non-distinguishable broadband 
component when speech is present (Figure \ref{fig:old_vs_new}.A-B). This challenge motivates our three main contributions: the introduction of a contrastive loss, a pre-trained deep speech representation, and a dedicated brain decoder.

\subsection{Model}
\subsubsection{Contrastive loss}
\label{sec:clip}
We reasoned that regression may be an ineffective loss because it departs from our objective \jr{ – \emph{i.e.} which requires maximally distinguishing different speech segments apart. Indeed, a regression objective stems from the principle that all of the dimensions of the Mel spectrogram are (1) equally important and (2) are scaled appropriately: the L2 objective inclines the model to predict low and high frequencies equally well, even if (1) some frequencies (e.g. very low)  may be irrelevant to speech and (2) some frequencies may vary in orders of magnitudes lowers than others.}{: decoding speech from brain activity.}
\jr{To relax this constraint, we opted for a contrastive objective and thus replaced the regression loss with}{Consequently, we replaced it with a contrastive loss, namely,} the ``CLIP'' loss (originally for Contrastive Language-Image Pre-Training) by \citet{radford2021learning}, which was originally designed to match latent representations in two modalities, text and images. \jr{Unlike the regression objective, this contrastive loss leads the model to find a combination of features that maximally discriminates samples in the batch. Consequently, the model is naturally inclined to focus on the informative dimensions of the Mel spectrograms, and to scale them appropriately.}{} 
We implement the CLIP loss as follows: Let $X$ be a brain recording segment and $Y\in \reel^{F\times T}$ the latent representation of its corresponding sound (a.k.a ``\emph{positive} sample''). We sample 
$N - 1 $
\emph{negative} samples $\bar{Y}_{j\in \{1, \ldots, N-1\}}$ over our dataset and we add the positive sample as $\bar{Y}_N = Y$.
We want our model to predict the probabilities 
$
\forall j \in\{1, \ldots, N\}, p_j = \proba{\bar{Y_j} = Y}.
$
We thus train a model $\enc$ mapping  the brain activity $X$ to a latent representation
$Z = \enc(X) \in \reel^{F\times T}$. The estimated probability can then be approximated by the dot product of $Z$ and the candidate speech latent representations $Y_j$, followed by a softmax:
\begin{linenomath}
\begin{equation}
\label{eq:clip_prob}
\hat{p}_{j} = \frac{\ee^{\dotprod{Z}{\bar{Y}_j}}}{\sum_{j'=1}^N \ee^{
    \dotprod{Z}{\bar{Y}_{j'}}}},
\end{equation}
with $\dotprod{\cdot}{\cdot}$ the inner product over both dimensions of $Z$ and $\hat{Y}$.
We then train $\enc$ with a cross-entropy between $p_{j}$ and $\hat{p}_{j}$. Note that for a large enough dataset,
we can neglect the probability of sampling twice the same segment, so that we have $p_j = \mathbbm{1}_{j = N}$,
and the cross-entropy simplifies to
\begin{equation}
\label{eq:clip_loss}
    L_\mathrm{CLIP}(p, \hat{p}) = -\log(\hat{p}_{N}) = 
        -\dotprod{Z}{Y} + 
        \log\Big(\sum_{j'=1}^{N}\ee^{\dotprod{Z}{ \bar{Y}_j'}} \Big).
\end{equation}
\end{linenomath}
Following \citep{radford2021learning}, we use the other elements of the batch as negative samples at train time. At test time, the negative samples correspond to all of the segments of the test \jr{set}{} but the positive one.

\subsubsection{Brain module}
\label{sec:brain_encoder}
\jr{For}{Finally, for} the brain module, we \jr{introduce}{use} a deep neural network $\enc$, input with raw M/EEG times series $X$ and a one-hot-encoding of the corresponding subject $s$, and outputs the latent brain representation $Z$, with the same sample rate as $X$.
This architecture consists of (1) a spatial attention layer over the M/EEG sensors followed (2) by a subject-specific 1x1 convolution designed to leverage inter-subject variability, which input to (3) a stack of convolutional blocks.
An overview of the model is given in Figure~\ref{fig:old_vs_new}. In the following, given a tensor $U$, we note $U^{(i,\ldots)}$
access to specific entries in the tensor.

\textbf{Spatial attention.} 
The brain data is first remapped onto $D_1 = 270$ channels with a spatial attention layer based on the location of the sensors. 
The 3D sensor locations are first projected on a 2D plane obtained with the MNE-Python function \texttt{find\_layout}~\citep{gramfort2013meg}, which uses a device-dependent surface designed to preserve the channel distances. Their 2D positions are finally normalized to $[0, 1]$.
For each output channel, a function over $[0, 1]^2$ is learnt, parameterized in the Fourier space. The weights over the input sensors is then given by the softmax of the function evaluated at the sensor locations.
Formally, each input channel $i$ has a location $(x_i, y_i)$ and each output channel $j$ is attached a function $a_j$ over $[0, 1]^2$ parameterized in the Fourier space as $z_j \in \complex^{K \times K}$ with $K{=}32$ harmonics along each axis, \emph{i.e.}
\begin{linenomath}
\begin{equation}
    a_j(x, y) = \sum_{k=1}^K\sum_{l=1}^K \mathrm{Re}(z_j^{(k, l)}) \cos\left(2 \pi (k x + l y)\right)  +\mathrm{Im}(z_j^{(k, l)}) \sin\left(2 \pi (k x + l y)\right).
\end{equation}
\end{linenomath}
The output is given by a softmax attention based on the evaluation of $a_j$ at each input position $(x_i, y_i)$:
\begin{linenomath}
\begin{equation}
    \forall j\in\{1, \ldots, D_1\}, \mathrm{SA}(X)^{(j)} =  \frac{1}{\sum_{i=1}^{C} \ee^{a_j(x_i, y_i)}}\left(
    \sum_{i=1}^{C} \ee^{a_j(x_i, y_i)} X^{(i)}
    \right)
\end{equation}
\end{linenomath}
with $\mathrm{SA}$ the spatial attention.
In practice, as $a_j$ is periodic, we scale down $(x, y)$ to keep a margin of 0.1 on each side.
We then apply a spatial dropout by sampling a location $(x_\mathrm{drop}, y_\mathrm{drop})$
 and removing from the softmax each sensor that is within a distance of $d_\mathrm{drop}=0.2$ of the sampled location.
We then add a 1x1 convolution (i.e. with a kernel size of 1) without activation and with the same number $D_1$ of output channels.

\textbf{Subject Layer.} 
To leverage inter-subject variability, we learn a matrix $M_{s} \in \reel^{D_1, D_1}$
for each subject $s \in [S]$ and apply it after the spatial attention layer along the channel dimension. This is similar but more expressive than the subject embedding used by~\citet{chehab2021deep} for MEG encoding, and follows decade of research on subject alignment \citep{xu2012regularized,haxby2020hyperalignment}. %

\looseness=-1
\textbf{Residual dilated convolutions.}
We then apply a stack of five blocks of three convolutional layers.
For the $k$-th block, the first two convolutions are applied with residual skip connections (except for the very first one where the number
of dimension potentially doesn't match), outputs $D_2=320$ channels and are followed by batch normalization~\citep{batchnorm} and a GELU activation~\citep{gelu}. 
The two convolutions are also dilated to increase their receptive field, respectively by $2^{2k\, \mathrm{mod} \,5}$  and 
$2^{2k + 1\, \mathrm{mod} \,5}$ (with $k$ zero indexed).
The third layer in a block outputs $2 D_2$ channels and uses a GLU activation~\citep{glu} which halves the number of channels.
All convolutions use a kernel  size of 3 over the time axis, a stride of 1, and sufficient padding to keep
the number of time steps constant across layers. 
The output of the model is obtained by applying two 
final 1x1 convolutions: first with $2 D_2$ outputs, followed by a GELU, and finally with $F$ channels as output, thus matching the dimensionality of speech representations. %
Given the expected delay between a stimulus and its corresponding brain responses, we further shift the input brain signal by 150\,ms into the future to facilitate the alignment between $Y$ and $Z$.

\subsubsection{Speech module}

The Mel spectrogram is a low-level representation of speech \jr{inspired from the cochlea}{} and is thus unlikely to match the rich variety of cortical representations \citep{hickok2007cortical}. Consequently, we replaced the Mel spectrograms with latent representations of speech. \jr{For this, we propose to either learn these representations}{that are either learned} end-to-end (``Deep Mel'' model) or \jr{to rely on those learnt by}{learned with} an independent self-supervised speech model (``wav2vec 2.0'', \citet{baevski2020wav2vec}).

\jr{\textbf{End-to-end speech representations with Deep Mel.}}{} 
\jr{The}{As detailed in the result section, the} ``Deep Mel'' module uses \jr{the same deep convolutional}{an} architecture \jr{}{similar }to the brain module \jr{devoid of the subject block, and thus simultaneously learns to extract speech and M/EEG representations such that they are maximally aligned. By definition, and unlike wav2vec 2.0, Deep Mel only sees the audio used in the M/EEG datasets. As this end-to-end approach proved to be less efficient than its pretrained counterpart based on wav2vec 2.0, we will thereafter focus on the latter.}{, but proved less efficient than its pretrained counterpart. We will thus focus the decoding results obtained with wav2vec 2.0.}

\jr{\textbf{Pretrained speech representations with wav2vec 2.0.}}{} Wav2vec 2.0 is trained \jr{with audio data only}{} to transform the raw waveform with convolutional and transformer blocks to predict masked parts of its own latent representations. \citet{baevski2020wav2vec} showed that the resulting model can be efficiently fine-tuned to achieve state-of-the-art performance in speech recognition.
Besides, this model effectively encodes a wide variety of linguistic features \citep{millet2022self,adolfi2022successes}. 
\jr{In particular}{Finally,} recent works show \jr{that the activations of wav2vec 2.0 linearly map onto those of the brain}{the existence of linear correspondence between the activations of the brain and those of wav2vec 2.0} \citep{millet2022toward,vaidya2022self}. Consequently, we here test whether this model effectively helps the present decoding task. In practice, we use the \texttt{wav2vec2-large-xlsr-53} \citep{ott2019fairseq}, which has been pre-trained on 56k hours of speech from 53 different languages.

\subsection{Datasets}
\label{sec:datasets}
 
\begin{table}
\centering
\caption{\textbf{Datasets}
}
\label{table:datasets}
\resizebox{\textwidth}{!}{
\begin{tabular}{lllrrr | rr | rr | r}

\toprule
     &&&&&& \multicolumn{2}{c}{\textit{Train set}}
     & \multicolumn{2}{c}{\textit{Test set}} \\
\textbf{Dataset} & \textbf{Lang.} & \textbf{Type} &  \textbf{Sensors} & \textbf{Subjects} & \textbf{Duration} & \textbf{Segments} & \textbf{Vocab.} & \textbf{Segments} & \textbf{Vocab.} & \jr{\textbf{Word overlap}}{} \\
\midrule
Broderick2019 & English & EEG & 128 & 19 & 19.2\,h& 2645 & 1418& 1842 & 764 & 67\%\\
Brennan2019 & English & EEG & 60 & 33 & 6.7\,h& 1211 & 513& 190 & 148 & 60\%\\
Schoffelen2019 & Dutch & MEG & 273 & 96 & 80.9\,h& 5497 & 1754& 1270 & 745 & 85\% \\
Gwilliams2022 & English & MEG & 208 & 27 & 56.2\,h& 4417 & 1810& 1363 & 846 & 64\%\\
\bottomrule
\end{tabular}
}
\end{table}
We test our approach on four public datasets, two based on MEG recordings and two on EEG. 
All datasets and their corresponding studies were approved by the relevant ethics committee and are publicly available for fundamental research purposes. Informed consent was obtained from each human research participant.
We provide an overview of the main characteristics of the datasets in Table~\ref{table:datasets}, including the number of train and test segments and vocabulary size over both splits. For all datasets, healthy adult volunteers passively listened to speech sounds (accompanied with some memory or comprehension questions to ensure participants were attentive), while their brain activity was recorded with MEG or EEG. 
In \citet{Schoffelen_19_204subjectMultimodalNeuroimaging}, Dutch-speaking participants listened to decontextualized Dutch sentences and word lists (Dutch sentences for which the words are randomly shuffled). The study was approved by the local ethics committee (CMO – the local “Committee on Research Involving Human Subjects” in the Arnhem-Nijmegen region). The data is publicly and freely available after registration on the Donders Repository. %
In \citet{gwilliams2020neural}, English-speaking participants listened to four fictional stories from the Masc corpus \citep{ide2010manually} in two identical sessions of one hour \citep{gwilliams2022masc}. The study was approved by the Institution Review Board (IRB) ethics committee of New York University Abu Dhabi.
In \citet{broderick2018electrophysiological}, English-speaking participants listened to extracts of ``The old man and the \jr{sea}{see}''. The study was approved by the Ethics Committees of the School of Psychology at Trinity College Dublin, and the Health Sciences Faculty at Trinity College Dublin.
In \citet{brennan2019hierarchical}, English-speaking participants listened to a chapter of ``Alice in Wonderland''. See Section~\ref{supp:datasets} in the Appendix for more details. The study was approved by the University of Michigan Health Sciences and Behavioral Sciences Institutional Review Board (HUM00081060).

\subsection{Preprocessing}
\label{sec:experiments}

M/EEG is generally considered to capture neural signals from relatively low frequency ranges \citep{hamalainen1993magnetoencephalography}. Consequently, we first resampled all brain recordings down to 120\,Hz with Torchaudio~\citep{yang2021torchaudio} and then split the data into \emph{training}, \emph{validation}, and \emph{testing} splits with a size roughly proportional to 70\%, 20\%, and 10\%.
We define a ``sample'' as a 3\,s window of brain recording with its associated speech representation.
A ``segment'' is a \emph{unique} 3\,s  window of speech sound. As the same segment can be presented to multiple subjects (or even within the same subject in \citet{gwilliams2020neural}), the splits are defined so that one segment is always assigned to the same split across repetitions. \jr{We ensure that there are no identical sentences across splits}{We ensure that there is no identical sentences across splits, and checked that each sentence was pronounced by a unique speaker}. Furthermore, we exclude all segments overlapping over different splits.
For clarity, we restrict the test segments to those that contain a word at a fixed location (here 500\,ms before word onset). 
 
M/EEG data can suffer from large artifacts, \emph{e.g.} eye movements, or variations in the 
electro-magnetic environment \citep{hamalainen1993magnetoencephalography}. To limit their impact, we apply a ``baseline correction'' (\emph{i.e.} we subtract to each input channel its average over the first 0.5\,s) and a robust scaler with scikit-learn \citep{pedregosa2011scikit}. We \jr{normalize the data and}{} clamp values greater than 20 \jr{standard deviations}{} to minimize the impact of large outlier samples. For the Mel spectrogram, we use 120 Mel bands (see Supplementary Section~\ref{sec:clamping})~\citep{young2002htk}, 
with a normalized STFT with a frame size of 512 samples and hop length of 128 samples, using audio sampled at 16k\,Hz. We apply
log-compression, \emph{i.e.} $\log(\epsilon + \mathrm{mel})$, with $\epsilon{=}10^{-5}$. When using wav2vec 2.0, we average
the activations of the last four layers of its transformer.
 We use standard normalization for both representations.

\subsection{Training}
\label{sec:training}

\vspace{-0.1cm}
One training epoch is defined as 1,200 updates using Adam~\citep{ADAM} with a learning rate of $3{\cdot}10^{-4}$
and a batch size of \jr{256}. We stop training when no improvement is observed on the valid set for 10 epochs and
keep the best model based on the valid loss. For the direct regression of the Mel spectrogram, we use the MSE loss.
We use two V100 GPUs with 16GB of memory.

\subsection{Evaluation}
\label{sec:evaluation}

\jr{\textbf{Mel reconstructions.} In Figure~\ref{fig:old_vs_new}, we illustrate some reconstructed Mel spectrograms using different models. With a regression loss, the generation of the Mel spectrogram is made directly. With a CLIP loss, we plot the weighted average across all test segments, where the weight corresponds to the probability of the segment to be true estimated with the CLIP loss. Specifically, given a segment and its matching audio (here the sentence ``Thank you for coming Ed''), we retrieve the predicted distribution over the  1,363 segments given by Eq. \eqref{eq:clip_prob}.}{} We then use this distribution to average the Mel spectrogram of each candidate segment.

\paragraph{Segment-level evaluation.}
The top-10 segment accuracy indicates whether the true segment is in the top-10 most likely segments \jr{predicted by the decoder. We favor reporting this metric over the standard top-1 accuracy, given the large number of possible segments as the model may be able to decode useful information, without necessarily guessing the exact speech segment.}{according to the probabilities.}

\paragraph{Word-level evaluation.} \jr{To}{We also} evaluate the model at the word level (\emph{e.g.} Figure \ref{fig:results_thankyou}), we select a 3\,s segment for each word of the test set \jr{(from -500\,ms to 2.5\,s)}{}. We input the model with the corresponding brain recordings, and output the probability distribution over all test segments. %
To obtain the distribution over the vocabulary, we group the candidate segments by \jr{the corresponding}{their first} word \jr{(\emph{i.e.} starting at t=0)}{} and sum the probabilities \jr{of the same words spoken in different segments}{within each group}. \jr{Top-1 and Top-10 word-level accuracy then quantify whether the true word is within the first or first ten most likely predictions of the model, respectively.}{}

\jr{\textbf{Prediction analysis.} To further inspect the predictions of the decoder, we quantify the extent to which they can be predicted from well-defined features $\tilde{Y} \in R^{n \times f_i}$. For this, we extract the phonetic features ($d=60$) with Phonemizer \citep{Bernard2021} the `zipf' frequency ($d=1$) with Wordfreq \citep{robyn_speer_2022_7199437}, the part-of-speech tags ($d=15$), the word embedding ($d=300$) of each word with spaCy \citep{spacy} as well as the phrase embedding of the corresponding 3\,s speech segment ($d=1,024$) with Laser \citep{schwenk2017learning}. We refer to `bag-of-words' the sum of word-embeddings over the segment. We then train a ridge regression with scikit-learn's default parameters \citep{pedregosa2011scikit} to predict the softmax probability of the true word output by the decoder, and estimate, with a five-split cross-validation, the correspondence between these two values with a Pearson $R$ correlation. In sum, this analysis quantifies how well the feature predicts the probability of being selected by the decoder.}{}

\jr{\textbf{Statistics.} Statistical comparison is performed on the test set. We use a Wilcoxon test across participants to compare different models on the same datasets. We use Mann-Whitney test across participants to compare different datasets.}{}

\subsection{Code availability}
The code to reproduce the present study is available at \jr{\href{https://github.com/facebookresearch/brainmagick}{github.com/facebookresearch/brainmagick}}{}.

\section{Results}

\begin{table}
\centering
\caption{\textbf{Results.} Top-10 segment-level accuracy (\%) for a random baseline model that predicts a uniform distribution over the segments (`random'), a convolutional network trained to predict the Mel spectrograms with a regression loss (`base'), the same model trained with a contrastive loss (`+ clip') and our model, \emph{i.e.} trained to predict the features of wav2vec 2.0 with a contrastive loss (`+ wav2vec 2.0').  $\pm$ indicates the standard deviation across three random initializations of the model's weights. }
\resizebox{\textwidth}{!}{
\begin{tabular}{lcccc}
\toprule
Model &     Brennan (EEG) &   Broderick (EEG) &   Gwilliams (MEG) &  Schoffelen (MEG) \\
\midrule
Random model  &          5.3$\pm$0.1 &          0.5$\pm$0.1 &          0.7$\pm$0.1 &          0.8$\pm$0.1 \\
Base Model    &          6.0$\pm$0.9 &          1.0$\pm$0.3 &         12.4$\pm$1.2 &         20.6$\pm$1.8 \\
+ CLIP        &          8.0$\pm$4.8 &          9.7$\pm$1.0 &         55.1$\pm$0.7 &         55.1$\pm$0.9 \\
+ Deep Mel    &         24.7$\pm$3.2 &         15.4$\pm$1.6 &         64.4$\pm$0.8 &         61.2$\pm$0.6 \\
+ wav2vec 2.0 &  \textbf{25.7}$\pm$2.9 &  \textbf{17.7}$\pm$0.6 &  \textbf{70.7}$\pm$0.1 &  \textbf{67.5}$\pm$0.4 \\
\bottomrule
\end{tabular}\label{tab:results}}

\end{table}

\subsection{Accurately decoding speech from M/EEG recordings}

Our model predicts the \jr{correct}{propper} segment, out of more than 1,000 \jr{possibilities}{possible ones}, with a top-10 accuracy \jr{up to 70.7\% on average across MEG subjects}{} (Table \ref{tab:results}, top-1 accuracy \jr{up to 41.3\%}{}, Table \ref{tab:results_top1}). For more than half of samples, the true audio segment is ranked first or second in the decoders' predictions. \jr{Interestingly, these performances can reach high top-1 accuracy in the best performing subjects: \emph{e.g.} above 80\% top-1 accuracy in the best participant of the dataset of \citet{gwilliams2022masc} (Figure \ref{fig:results}{}A).}{} For comparison, a model that predicts a uniform distribution over the vocabulary (`random model') only achieves \jr{less than 1\%}{a 2\%} top-10 accuracy on the same MEG datasets. Decoding performance for EEG datasets is lower: our model reaches \jr{17.7}{19}\% and \jr{25.7}{31}\% top-10 accuracy \jr{for the two EEG datasets presently analyzed}{}. While modest, these scores are \jr{much}{four times} higher than the random baseline. %

\subsection{Is MEG really much better than EEG?}

\jr{To investigate whether these performances depend on the total recording duration and/or the number of recording sensors, we train our model on a subset of the data which homogenize recording time, number of sensors, number of participants. For this, we discard the dataset from \citet{brennan2019hierarchical}, to avoid over-limiting this analysis datasets. Consequently, we match all datasets to the smallest number of channels of the three remaining datasets by keeping a random but fixed subset of channels (\emph{e.g.} 128). We keep only 19 subjects per datasets, again aligning on the smallest for all three datasets. Finally, we keep the same average duration per subject for all three datasets, by dropping out some training segments (\emph{i.e.} the same segments are dropped for all subjects or repetitions within one subject). All test  segments are kept to maximize reliability. Overall, this subsampling diminishes decoding performance (\emph{e.g.} top-10: 30.3\%  for \citet{Schoffelen_19_204subjectMultimodalNeuroimaging} and 31.7\% for \citet{gwilliams2020neural}), but MEG decoding remains much better than the EEG  (Mann-Whitney across MEG and EEG subjects: all $p<10^{-6}$ ). While these results should be confirmed by using the same stimuli to participants recorded either with EEG or MEG, they suggest that the difference of decoding performance observed between studies is mainly driven by the type of device.}{}

\subsection{`Speech module' evaluation.}

\jr{To evaluate our approach, we compare these decoding performances to those obtained with models that target different representations of speech (Figure \ref{fig:results} and Table \ref{tab:results}).}{}
\jr{While a model trained to predict the Mel spectrogram with a regression objective (`Base model' in Table \ref{tab:results}) is systematically higher than chance, the use of a contrastive loss (`+ CLIP') leads to decoding gains between that range from 2\% (for \citet{brennan2019hierarchical}) to 42.7\% (for \citet{gwilliams2020neural}). This gain is further supplemented by targeting latent representation of the Mel spectrogram (`+ Deep Mel'). The latent representations of speech sounds, however, appear to be best identified with a pretrained speech module, \emph{i.e.} by using `wav2vec 2.0', a model pretrained with self-supervised learning on speech sounds only, rather than by jointly learning speech and M/EEG representations (Table~\ref{tab:results}). Overall, these results show the importance, for decoding, to target latent representations of speech.}{}

\subsection{`Brain module' evaluation.}
\jr{To evaluate the elements of the brain module, we performed a series of ablation experiments, and trained the corresponding models on the same data. Overall, several elements}{ Finally, other modestly but design choices significantly} impact performance. Performance systematically decreases when removing skip connections, the spatial attention module, the initial or final convolutional layers (Table \ref{tab:ablation}). \jr{These results}{We} also show how essential clamping is to train the model.
Additional experiments confirm that the present end-to-end architecture is robust to M/EEG artefacts, and thus requires little preprocessing (Supplementary sections~\ref{sec:autoreject} and \ref{sec:clamping}).

\begin{figure}[ht]
  \centering
  \includegraphics[width=\textwidth]{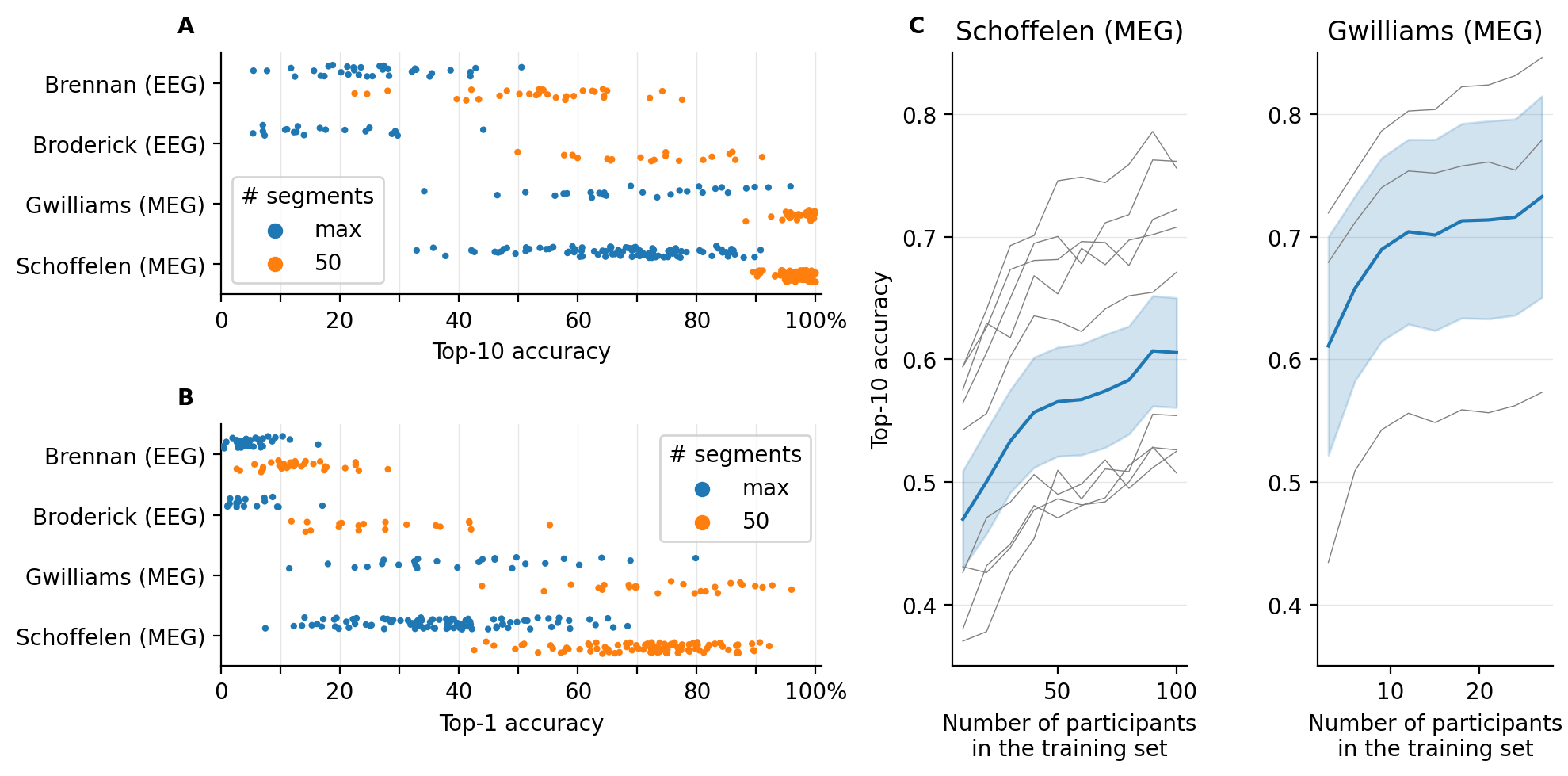}
  \caption{
  \red
  \textbf{A.} Each dot represents the top-10 accuracy of a single subject, as estimated either with the full test set (blue) or with 50 possible segments (orange). \textbf{B.} Same as A, for top-1 accuracy.
  \textbf{C.} \jrr{Top-10 accuracy as a function of the number of participants in the training set (blue line) as evaluated on the first 10\% of the participants. The error bar indicates the standard error of the mean (SEM) across participants (gray lines)}{}.
  \color{black}
  }
  \label{fig:results}
\end{figure}

\jr{Importantly, these ablations analyses also reveal the importance of the subject layer. Note that this gain is modest when compared to performance obtained with the subject embedding we introduced recently~\citep{chehab2021deep}. To further investigate}{Third, to test }
whether our model effectively leverage the inter-individual variability,
we trained it on a variable number of subjects and computed its accuracy on the first 10\% of subjects. As shown in Figure \ref{fig:results}C, decoding performance \jr{steadily}{} increases as the model is trained with more subjects on the two MEG datasets.

\begin{table}[]
\begin{center}
\caption{\textbf{Ablations.} Top-10 segment-level accuracy (\%) for our model and its ablated versions. \jrr{Stars indicate significant gain ($p<0.001$) across participants. Confidence intervals are computed as Standard Error of the Mean (SEM) over 3 runs.}{}}
\resizebox{\textwidth}{!}{

\begin{tabular}{lcccc | c | c}
\toprule
 &   Broderick (EEG) &     Brennan (EEG) &  Schoffelen (MEG) &   Gwilliams (MEG) &   delta &        p-val \\
Model                       &                   &                   &                   &                   &         &              \\
\midrule
Our model                   &  \textbf{17.7} $\pm$ 0.6 &         25.7 $\pm$ 2.9 &  \textbf{67.5} $\pm$ 0.4 &  \textbf{70.7} $\pm$ 0.1 &         &              \\
- Spatial attention dropout &       16.0 $\pm$ 1.7 * &  \textbf{26.8} $\pm$ 0.7 &         67.5 $\pm$ 0.2 &       69.0 $\pm$ 0.2 * &   $0.4$ &        0.009 \\
- GELU + ReLU               &         16.4 $\pm$ 0.1 &         24.6 $\pm$ 2.1 &       65.8 $\pm$ 0.6 * &       68.8 $\pm$ 1.3 * &   $1.6$ &  $<10^{-18}$ \\
- Final convs               &       14.2 $\pm$ 1.1 * &       19.0 $\pm$ 4.4 * &         67.5 $\pm$ 0.3 &       68.9 $\pm$ 0.9 * &   $1.1$ &  $<10^{-10}$ \\
- Non-residual GLU conv     &        8.4 $\pm$ 6.8 * &        6.0 $\pm$ 0.2 * &         67.0 $\pm$ 0.2 &         70.2 $\pm$ 0.2 &   $1.6$ &  $<10^{-10}$ \\
- Skip connections          &       13.9 $\pm$ 2.0 * &         24.2 $\pm$ 2.7 &       65.4 $\pm$ 0.4 * &       66.2 $\pm$ 0.3 * &   $2.4$ &  $<10^{-21}$ \\
- Initial 1x1 conv          &         15.4 $\pm$ 0.6 &       22.1 $\pm$ 1.9 * &       62.9 $\pm$ 0.9 * &       67.7 $\pm$ 0.7 * &   $3.4$ &  $<10^{-26}$ \\
- Spatial attention         &       15.4 $\pm$ 0.6 * &       20.6 $\pm$ 2.2 * &       65.9 $\pm$ 0.3 * &       65.5 $\pm$ 0.4 * &   $2.5$ &  $<10^{-22}$ \\
- Subj layer                &        8.1 $\pm$ 1.9 * &       20.2 $\pm$ 1.3 * &       42.4 $\pm$ 0.1 * &       47.0 $\pm$ 1.3 * &  $14.4$ &  $<10^{-28}$ \\
- Clamping                  &        0.5 $\pm$ 0.0 * &       14.1 $\pm$ 1.0 * &        1.5 $\pm$ 0.3 * &      23.6 $\pm$ 24.6 * &  $26.2$ &  $<10^{-29}$ \\
\bottomrule
\end{tabular}

}\label{tab:ablation}

\end{center}
\end{table}

\subsection{Decoded representations best correlate with phrase embeddings.}
\label{section:errors}

\begin{figure}[ht]
  \centering
  \includegraphics[width=\textwidth]{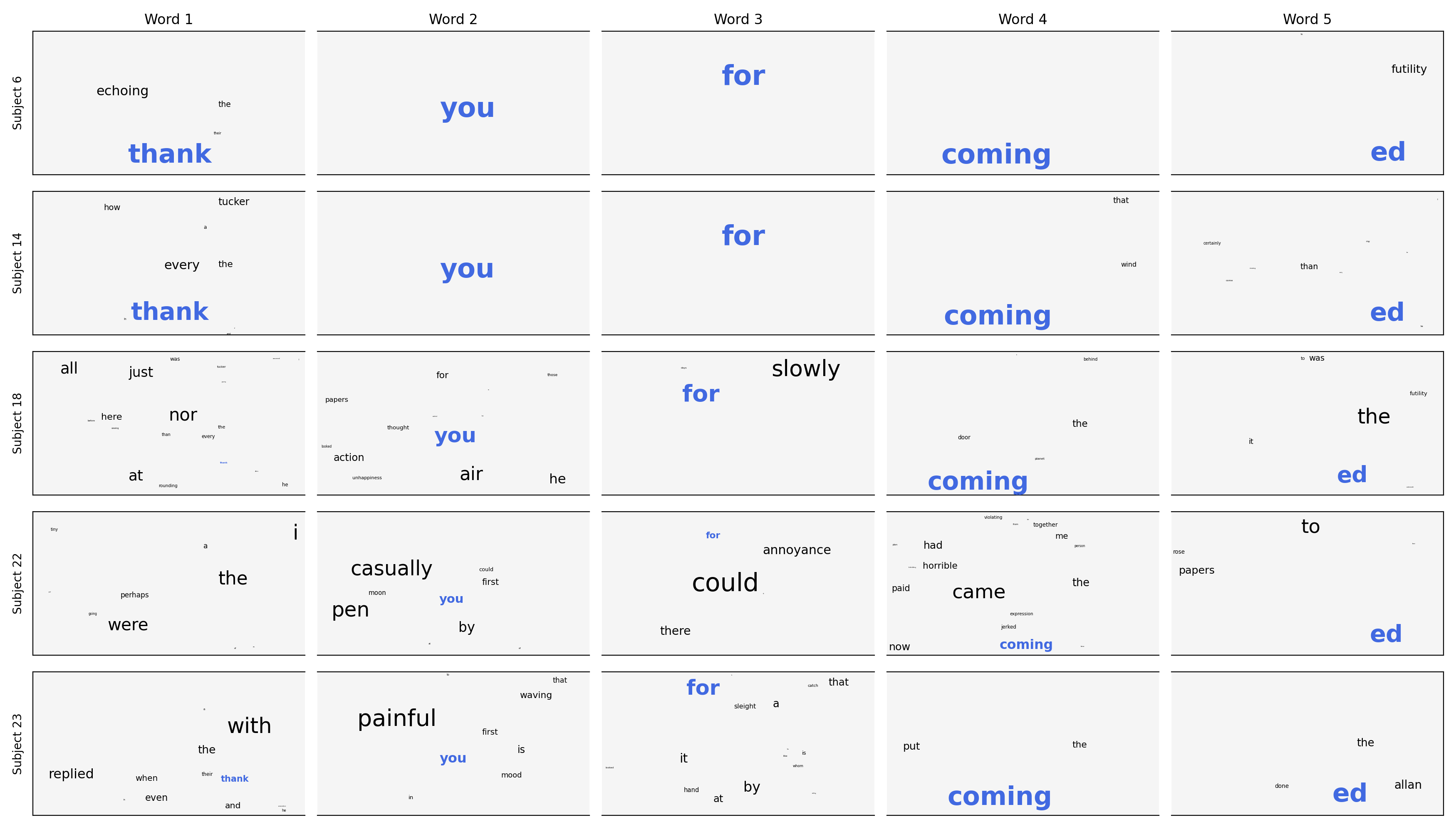}
  \caption{
    Word-level predictions for \jr{five}{three} representative subjects \jr{(between the 20\% (top) and the 80\% percentiles (bottom) of the cohort)}{} of \citet{gwilliams2020neural} \jr{while they listened}{listening} to the sentence ``Thank you for coming, Ed''. \jrr{Blue words correspond to the correct word, black words to negative candidates}{}. Text size is proportional to the log-probability output by our model.
}

  \label{fig:results_thankyou}
\end{figure}

\begin{figure}[ht]
  \centering
  \includegraphics[width=\textwidth]{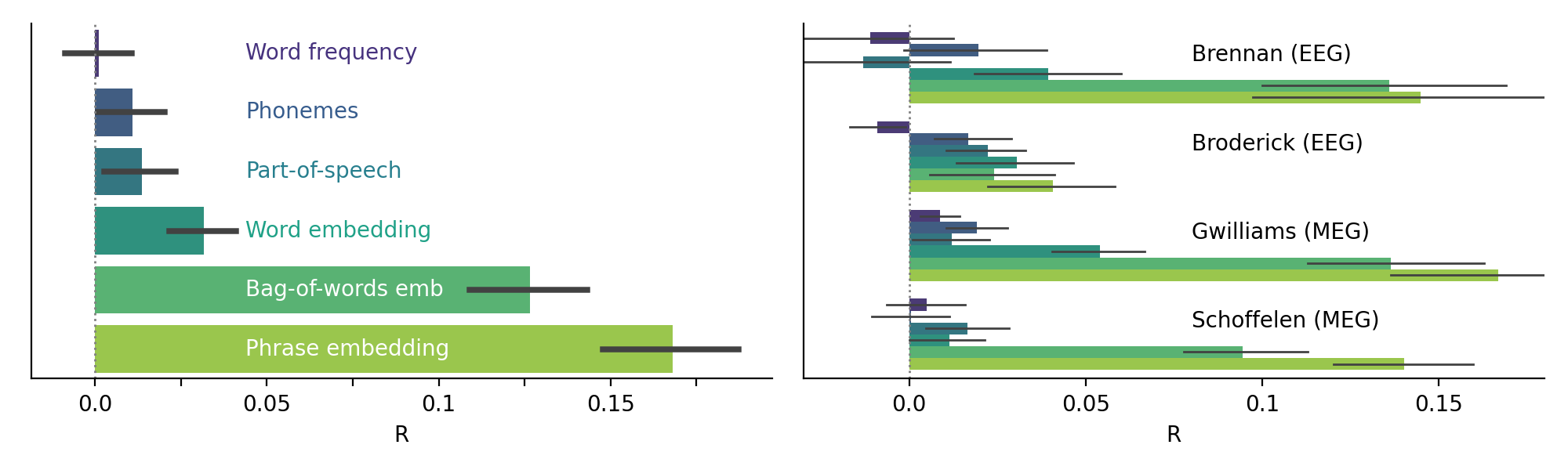}
  \caption{
  \jr{The $R$ values quantify the extent to which phonemes, word frequency, part-of-speech, word embedding, and phrase embedding respectively predict the probability of the predicted word to be correct}{}. Error bars are the SEM across participants.
  \label{fig:errors}
}
\end{figure}

\jr{What type of representations does our model use to decode speech from brain signals? This interpretability question is notoriously difficult to address \citep{angrick2019interpretation,king2018encoding}. In an attempt to nonetheless shed light on this issue, we analyze the single-word and single-segment predictions of our model. Figure \ref{fig:results_thankyou} and Supplementary Figure \ref{fig:results_sentences} illustrate such predictions: \emph{i.e.} the probability of each word given the MEG data of five representative subjects, and five representative segments of \citet{gwilliams2020neural}, respectively. Then, we train a linear regression to predict the softmax probability of the true word estimated by the decoder, given different set of features, ranging from low-level (\emph{e.g.} phonemes) to high-level representations (\emph{e.g.} phrase embedding, see Methods for details). The results, displayed in Figure \ref{fig:errors}, show that the part-of-speech ($p<0.004$), word embedding ($p<10^{-8}$), bag-of-words ($p<10^{-23}$) and phrase embedding ($p<10^{-23}$) significantly predict the single-trial decoding predictions. Given that phrase embeddings are known to capture semantic and syntactic representations \citep{hewitt2019structural,caucheteux2020language,caucheteux2022deep}, these correlations suggest that our model decodes relatively high-level representations of speech.}{}

\red
\section{Discussion}
\label{sec:discussion}

Overall, our model accurately identifies, from 3 seconds of non-invasive recordings of brain activity, the corresponding speech segment with up to 41\% accuracy out of more than 1,000 distinct possibilities. This performance, sometimes reaching 80\% in the very best participants allows the decoding of perceived \jrr{words and}{} phrases absent from the training set.

To decode speech perception from M/EEG, two major challenges must be overcome. First, these signals can be very noisy, making it difficult to extract useful information. Second, it is unclear which features of speech are, in fact, represented in the brain. In this section, we will discuss how our ``brain'' and ``speech'' modules respectively address these two issues in the case of speech \emph{perception}. Finally, we evaluate the performance of our model in comparison to previous works and outline the necessary steps to be taken before hoping to deploy this approach for the decoding of speech \emph{production} in clinical settings.

\subsection{Efficiently extracting brain signals.}

Non-invasive recordings are notoriously noisy: these signals present large variations across trials and subjects and they are often contaminated by large artifacts \citep{hamalainen1993magnetoencephalography,schirrmeister2017deep,king2018encoding}. Historically, solving this problem has involved complex preprocessing pipelines that include a variety of techniques – independent component analysis, outlier detection, artifact correction – that end by a linear model specific to each participant \citep{haxby2001distributed,kamitani2005decoding,nishimoto2011reconstructing}. More recently, several deep learning architectures have proved successful in solving simple classification tasks trained on single-subject recordings \citep{roy2019deep,dash2018determining}. 

Building on these efforts, our end-to-end architecture requires minimal preprocessing of M/EEG signals and can be trained with a variety of participants, devices, and stimuli. As decoding speech production can be challenged by the presence of muscular activity, we here evaluate this model on four public datasets where healthy participants \emph{listened} to natural speech. 
Our analyses suggest that advanced M/EEG preprocessing does not provide a major advantage in the current decoding task and that a simple baseline correction followed by a robust scaler and clamping suffices. In addition, not only does our subject-specific layer improves decoding performance, but this performance increases with the amount of participants present in the training set. These findings, combined with both the rise of publicly available datasets and the potential to learn informative features from unannotated data \citep{banville2021uncovering,thomas2022self} suggest that this brain module may be a stepping stone for building a foundational model of brain recordings.

\subsection{How is language represented in the brain?}

Separating noise and signal in brain recordings is not, however, the only challenge. The nature of these representations in terms of their acoustic, phonetic, lexical, and semantic properties remains poorly known. Consequently, determining the representations most suitable for decoding is an unresolved problem. To tackle this issue, previous studies have primarily used supervised models targeting well-defined features of language, such as individual letters, phonemes, or frequency bands of the audio spectrogram \citep{miyawaki2008visual,pasley2012reconstructing,dash2018determining,jayaram2018moabb,lawhern2018eegnet,jahangiri2018covert,panachakel2021decoding,willett2021high,orpella2022speech,ali2022enhancing}. Although this approach has demonstrated clear successes, it may impede the speed at which words are decoded from brain activity: For instance, spelling out a word letter by letter could be a slow and laborious process. As an alternative, others have proposed to learn to classify a small set of words \cite{chan2011decoding,koizumi2018development,garcia2019transfer,dash2019decoding,murphy2022decoding}. This approach, however, is difficult to scale to a vocabulary size adequate for natural language. Finally, word semantics may be directly decoded from fMRI signals \citep{horikawa2017generic,gauthier2019linking,affolter2020brain2word,pascual2022improving,fernandino2022decoding,tang2022semantic}. However, the corresponding performances remain currently modest at the single trial level.

Here, we show how a neural network pretrained on a large corpus of speech sounds provides representations of language that are particularly valuable for brain decoding. Specifically, we leverage the recent discovery that these self-supervised speech models learn features that linearly relate to those of the brain \citep{millet2022toward,vaidya2022self} to build our speech module. By applying contrastive learning, we can effectively identify the most appropriate features for identifying new speech segments. Our analyses confirm that this approach outperforms (1) a supervised decoding of the MEL spectrogram as well as (2) `Deep Mel`, \emph{i.e.} latent representations of the MEL spectrogram optimized for decoding solely from the present M/EEG datasets. Finally, the inspection of the decoding predictions suggests that our model primarily captures the lexical and contextual features captured by modern word embeddings and language models. To date, however, what these high-level features represent and how these representations are structured and manipulated remain to be determined.

\subsection{Comparison to previous works}

Comparing the performance of our model to previous works is difficult because the variety of experimental paradigms is not compensated by a profusion of open datasets and reproducible code. Two elements may, nonetheless, substantiate such a comparison.

First, the size of vocabulary presently considered exceeds previous attempts, often by several orders of magnitude: For example, M/EEG studies typically used supervised decoders to discriminate a very small set of words \cite{chan2011decoding,koizumi2018development,garcia2019transfer,dash2019decoding,murphy2022decoding} or sublexical classes (e.g. phonemes, syllables, tones) \citep{jayaram2018moabb,lawhern2018eegnet,jahangiri2018covert,panachakel2021decoding,orpella2022speech,ali2022enhancing}. For example, \citet{sun2016neural}, \citet{sree2017vowel}, and \citet{moinnereau2018classification} all developed a decoder to classify 11, 5 and 2 distinct imagined phonemes, respectively, from EEG signals. Similarly, \citet{lopopolo2020part,chan2011decoding,nguyen2017inferring} respectively developed an MEG decoder to classify 6 distinct part-of-speech (with 48\% accuracy), 10 words (83\% accuracy) and 3 words (70\% accuracy). The limited vocabulary used in these non-invasive studies contrast with the present approach, which demonstrably accurately distinguishes several hundreds of words. Furthermore, the performances of our model are based on vocabularies which do not fully overlap with those used in the training set (Table \ref{table:datasets}). For example, for the Gwilliams dataset, decoding performance reaches 40\% in spite of the fact that nearly 36\% of the words were never presented during training. Overall, such zero-shot decoding shows the versatility of this approach and opens the possibility to decode from even larger vocabulary.

Second, although our model's performance remains modest, it may not be too distant from the performance obtained with \emph{invasive} recordings of brain activity. Indeed, decoding the perception of \emph{isolated words} from a vocabulary of n=50 words leads to a top-1 accuracy of 22.7\% on average, but up to 42.9\% in the best participants (Supplementary Figure \ref{fig:single}) In comparison, \cite{moses2021neuroprosthesis} report decoding \emph{produced} words from intracranial recordings with a top-1 accuracy of 39.5\% for isolated words out of n=50 words. Similarly, still restricting the number of candidates to 50 and, this time, within the context of a \emph{sentence}, our model decoding is above 72.5\% top-1 accuracy on average across participants, and the very best participants reach between 92.2\% \citep{Schoffelen_19_204subjectMultimodalNeuroimaging} and 95.9\% (\citep{gwilliams2022masc}, Figure \ref{fig:results}B), where \citet{moses2021neuroprosthesis} reached a top-1 accuracy of 74\%. While comparing the decoding of perceived \emph{vs} produced words should be considered with caution given their different brain bases, the performance of the current model thus leads us to be optimistic about its potential applicability in a speech production context.

\subsection{Remaining steps to decode speech production in the clinics}

Our non-invasive approach focuses on speech \emph{perception}. To reach the performance obtained with clinical recordings \citep{herff2015brain,martin2016word,angrick2019speech,willett2021high,moses2021neuroprosthesis,angrick2021real,kohler2021synthesizing,metzger2022generalizable}, decoding intended communication will thus require addressing several challenges. Three specific challenges stand out.

First, the current model needs to be adapted to speech \emph{production}. This can, in principle, be achieved by replacing the speech module with a neural network pre-trained on production tasks such as handwriting or speech production.

Second, the current contrastive learning objective can only identify the most likely word or speech segment from a predetermined set. The model thus needs to be supplemented with a generative module that can estimate the most likely phoneme, word, or sentence given brain activity without requiring this set of candidates, similarly to what is being achieved with fMRI \citep{tang2022semantic,ozcelik2023brain}.

Finally, our study reveals striking differences between EEG and MEG. While EEG is known to be less precise than MEG, we did not expect such an important difference of decoding performance. Adapting current MEG systems to the clinics will require substantial efforts, however: while new room-temperature sensors already show signal-to-noise ratio comparable to the superconducting quantum interference devices (SQUIDs) used in the present study, these systems are not commonly deployed in clinical setting, whose magnetic environment can be extremely noisy. Combined with A.I. systems, these new devices could nevertheless contribute to improve the diagnosis, prognosis and restoration of language processing in non- or poorly-communicating patients without putting them at risk for brain surgery. In that regard, we hope that the release of a fully-reproducible pipeline will contribute to the development of safe and scalable non-invasive methods for decoding intended communication.
\color{black}

\section{Data Availability}

The data from \citet{Schoffelen_19_204subjectMultimodalNeuroimaging}
 was provided (in part) by the Donders Institute for Brain, Cognition and Behaviour with a ``RU-DI-HD-1.0'' licence.%
 The data for \citet{gwilliams2020neural} is available under
 CC0 1.0 Universal. %
 The data for \citet{broderick2018electrophysiological}
 is available under the same licence. %
 Finally, the data from \citet{brennan2019hierarchical}
 is available under the CC BY 4.0 licence. %
All audio files were provided by the authors of each dataset. 

\section{Code Availability}

The complete source code for processing the datasets, training and evaluating the models and method presented here
are available at \href{https://github.com/facebookresearch/brainmagick}{github.com/facebookresearch/brainmagick}. The 
code is provided under the CC-NC-BY 4.0 license.
We also provide a fixed version of the code as of the release of the present articel,
on Zenodo under the DOI \href{https://zenodo.org/record/8114374
}{10.5281/zenodo.8114374}

\section{Acknowledgments}
This work was funded in part by FrontCog grant ANR-17-EURE-0017 to JRK for his work at PSL.

\section{Statement of contributions.} 

The project was led by Alexandre Défossez and Jean-Remi King. Jean-Remi King took care of the data curation.
The training pipeline was built by Jérémy Rapin, Ori Kabeli and Alexandre Défossez. Ori Kabeli and Alexandre Défossez 
handled model training and hyper-parameter search.
The speech module and evaluation pipeline was built by Charlotte Caucheteux.
Alexandre Défossez, Charlotte Caucheteux, Ori Kabeli and Jean Remi King provided in depth data and results analysis.
The present paper was written by Alexandre Défossez, Charlotte Caucheteux, and Jean-Remi King.

\section{Competing interests}
The authors declare no competing interests.

\clearpage

\bibliography{references}

\begin{thebibliography}{96}
\providecommand{\natexlab}[1]{#1}
\providecommand{\url}[1]{\texttt{#1}}
\expandafter\ifx\csname urlstyle\endcsname\relax
  \providecommand{\doi}[1]{doi: #1}\else
  \providecommand{\doi}{doi: \begingroup \urlstyle{rm}\Url}\fi

\bibitem[Adolfi et~al.(2022)Adolfi, Bowers, and Poeppel]{adolfi2022successes}
Federico Adolfi, Jeffrey~S Bowers, and David Poeppel.
\newblock Successes and critical failures of neural networks in capturing
  human-like speech recognition.
\newblock \emph{arXiv preprint arXiv:2204.03740}, 2022.

\bibitem[Affolter et~al.(2020)Affolter, Egressy, Pascual, and
  Wattenhofer]{affolter2020brain2word}
Nicolas Affolter, Beni Egressy, Damian Pascual, and Roger Wattenhofer.
\newblock Brain2word: decoding brain activity for language generation.
\newblock \emph{arXiv preprint arXiv:2009.04765}, 2020.

\bibitem[AI(2017)]{spacy}
Explosion AI.
\newblock spacy.
\newblock 2017.
\newblock URL \url{https://spacy.io/}.

\bibitem[Akbari et~al.(2019)Akbari, Khalighinejad, Herrero, Mehta, and
  Mesgarani]{akbari2019towards}
Hassan Akbari, Bahar Khalighinejad, Jose~L Herrero, Ashesh~D Mehta, and Nima
  Mesgarani.
\newblock Towards reconstructing intelligible speech from the human auditory
  cortex.
\newblock \emph{Scientific reports}, 9\penalty0 (1):\penalty0 1--12, 2019.

\bibitem[Ali et~al.(2022)Ali, Saif-ur Rehman, Dyck, Glasmachers, Iossifidis,
  and Klaes]{ali2022enhancing}
Omair Ali, Muhammad Saif-ur Rehman, Susanne Dyck, Tobias Glasmachers, Ioannis
  Iossifidis, and Christian Klaes.
\newblock Enhancing the decoding accuracy of eeg signals by the introduction of
  anchored-stft and adversarial data augmentation method.
\newblock \emph{Scientific reports}, 12\penalty0 (1):\penalty0 1--19, 2022.

\bibitem[Angrick et~al.(2019{\natexlab{a}})Angrick, Herff, Johnson, Shih,
  Krusienski, and Schultz]{angrick2019interpretation}
Miguel Angrick, Christian Herff, Garett Johnson, Jerry Shih, Dean Krusienski,
  and Tanja Schultz.
\newblock Interpretation of convolutional neural networks for speech
  spectrogram regression from intracranial recordings.
\newblock \emph{Neurocomputing}, 342:\penalty0 145--151, 2019{\natexlab{a}}.

\bibitem[Angrick et~al.(2019{\natexlab{b}})Angrick, Herff, Mugler, Tate,
  Slutzky, Krusienski, and Schultz]{angrick2019speech}
Miguel Angrick, Christian Herff, Emily Mugler, Matthew~C Tate, Marc~W Slutzky,
  Dean~J Krusienski, and Tanja Schultz.
\newblock Speech synthesis from ecog using densely connected 3d convolutional
  neural networks.
\newblock \emph{Journal of neural engineering}, 16\penalty0 (3):\penalty0
  036019, 2019{\natexlab{b}}.

\bibitem[Angrick et~al.(2021)Angrick, Ottenhoff, Diener, Ivucic, Ivucic,
  Goulis, Saal, Colon, Wagner, Krusienski, et~al.]{angrick2021real}
Miguel Angrick, Maarten~C Ottenhoff, Lorenz Diener, Darius Ivucic, Gabriel
  Ivucic, Sophocles Goulis, Jeremy Saal, Albert~J Colon, Louis Wagner, Dean~J
  Krusienski, et~al.
\newblock Real-time synthesis of imagined speech processes from minimally
  invasive recordings of neural activity.
\newblock \emph{Communications biology}, 4\penalty0 (1):\penalty0 1--10, 2021.

\bibitem[Anumanchipalli et~al.(2019)Anumanchipalli, Chartier, and
  Chang]{anumanchipalli2019speech}
Gopala~K Anumanchipalli, Josh Chartier, and Edward~F Chang.
\newblock Speech synthesis from neural decoding of spoken sentences.
\newblock \emph{Nature}, 568\penalty0 (7753):\penalty0 493--498, 2019.

\bibitem[Baevski et~al.(2020)Baevski, Zhou, Mohamed, and
  Auli]{baevski2020wav2vec}
Alexei Baevski, Yuhao Zhou, Abdelrahman Mohamed, and Michael Auli.
\newblock wav2vec 2.0: A framework for self-supervised learning of speech
  representations.
\newblock \emph{Advances in Neural Information Processing Systems},
  33:\penalty0 12449--12460, 2020.

\bibitem[Banville et~al.(2021)Banville, Chehab, Hyv{\"a}rinen, Engemann, and
  Gramfort]{banville2021uncovering}
Hubert Banville, Omar Chehab, Aapo Hyv{\"a}rinen, Denis-Alexander Engemann, and
  Alexandre Gramfort.
\newblock Uncovering the structure of clinical eeg signals with self-supervised
  learning.
\newblock \emph{Journal of Neural Engineering}, 18\penalty0 (4):\penalty0
  046020, 2021.

\bibitem[Bernard and Titeux(2021)]{Bernard2021}
Mathieu Bernard and Hadrien Titeux.
\newblock Phonemizer: Text to phones transcription for multiple languages in
  python.
\newblock \emph{Journal of Open Source Software}, 6\penalty0 (68):\penalty0
  3958, 2021.
\newblock \doi{10.21105/joss.03958}.
\newblock URL \url{https://github.com/bootphon/phonemizer}.

\bibitem[Birbaumer et~al.(1999)Birbaumer, Ghanayim, Hinterberger, Iversen,
  Kotchoubey, K{\"u}bler, Perelmouter, Taub, and Flor]{birbaumer1999spelling}
Niels Birbaumer, Nimr Ghanayim, Thilo Hinterberger, Iver Iversen, Boris
  Kotchoubey, Andrea K{\"u}bler, Juri Perelmouter, Edward Taub, and Herta Flor.
\newblock A spelling device for the paralysed.
\newblock \emph{Nature}, 398\penalty0 (6725):\penalty0 297--298, 1999.

\bibitem[Boto et~al.(2018)Boto, Holmes, Leggett, Roberts, Shah, Meyer,
  Mu{\~n}oz, Mullinger, Tierney, Bestmann, et~al.]{boto2018moving}
Elena Boto, Niall Holmes, James Leggett, Gillian Roberts, Vishal Shah, Sofie~S
  Meyer, Leonardo~Duque Mu{\~n}oz, Karen~J Mullinger, Tim~M Tierney, Sven
  Bestmann, et~al.
\newblock Moving magnetoencephalography towards real-world applications with a
  wearable system.
\newblock \emph{Nature}, 555\penalty0 (7698):\penalty0 657--661, 2018.

\bibitem[Brennan and Hale(2019)]{brennan2019hierarchical}
Jonathan~R Brennan and John~T Hale.
\newblock Hierarchical structure guides rapid linguistic predictions during
  naturalistic listening.
\newblock \emph{PloS one}, 14\penalty0 (1):\penalty0 e0207741, 2019.

\bibitem[Broderick et~al.(2018)Broderick, Anderson, Di~Liberto, Crosse, and
  Lalor]{broderick2018electrophysiological}
Michael~P Broderick, Andrew~J Anderson, Giovanni~M Di~Liberto, Michael~J
  Crosse, and Edmund~C Lalor.
\newblock Electrophysiological correlates of semantic dissimilarity reflect the
  comprehension of natural, narrative speech.
\newblock \emph{Current Biology}, 28\penalty0 (5):\penalty0 803--809, 2018.

\bibitem[Brumberg et~al.(2009)Brumberg, Kennedy, and
  Guenther]{brumberg2009artificial}
Jonathan~S Brumberg, Philip~R Kennedy, and Frank~H Guenther.
\newblock Artificial speech synthesizer control by brain-computer interface.
\newblock In \emph{Tenth Annual Conference of the International Speech
  Communication Association}, 2009.

\bibitem[Caucheteux and King(2020)]{caucheteux2020language}
Charlotte Caucheteux and Jean-R{\'e}mi King.
\newblock Language processing in brains and deep neural networks: computational
  convergence and its limits.
\newblock \emph{BioRxiv}, 2020.

\bibitem[Caucheteux et~al.(2022)Caucheteux, Gramfort, and
  King]{caucheteux2022deep}
Charlotte Caucheteux, Alexandre Gramfort, and Jean-R{\'e}mi King.
\newblock Deep language algorithms predict semantic comprehension from brain
  activity.
\newblock \emph{Scientific Reports}, 12\penalty0 (1):\penalty0 16327, 2022.

\bibitem[Chan et~al.(2011)Chan, Halgren, Marinkovic, and
  Cash]{chan2011decoding}
Alexander~M Chan, Eric Halgren, Ksenija Marinkovic, and Sydney~S Cash.
\newblock Decoding word and category-specific spatiotemporal representations
  from meg and eeg.
\newblock \emph{Neuroimage}, 54\penalty0 (4):\penalty0 3028--3039, 2011.

\bibitem[Chehab et~al.(2021)Chehab, Defossez, Loiseau, Gramfort, and
  King]{chehab2021deep}
Omar Chehab, Alexandre Defossez, Jean-Christophe Loiseau, Alexandre Gramfort,
  and Jean-Remi King.
\newblock Deep recurrent encoder: A scalable end-to-end network to model brain
  signals.
\newblock \emph{arXiv preprint arXiv:2103.02339}, 2021.

\bibitem[Claassen et~al.(2019)Claassen, Doyle, Matory, Couch, Burger,
  Velazquez, Okonkwo, King, Park, Agarwal, et~al.]{claassen2019detection}
Jan Claassen, Kevin Doyle, Adu Matory, Caroline Couch, Kelly~M Burger, Angela
  Velazquez, Joshua~U Okonkwo, Jean-R{\'e}mi King, Soojin Park, Sachin Agarwal,
  et~al.
\newblock Detection of brain activation in unresponsive patients with acute
  brain injury.
\newblock \emph{New England Journal of Medicine}, 380\penalty0 (26):\penalty0
  2497--2505, 2019.

\bibitem[Cruse et~al.(2011)Cruse, Chennu, Chatelle, Bekinschtein,
  Fern{\'a}ndez-Espejo, Pickard, Laureys, and Owen]{cruse2011bedside}
Damian Cruse, Srivas Chennu, Camille Chatelle, Tristan~A Bekinschtein, Davinia
  Fern{\'a}ndez-Espejo, John~D Pickard, Steven Laureys, and Adrian~M Owen.
\newblock Bedside detection of awareness in the vegetative state: a cohort
  study.
\newblock \emph{The Lancet}, 378\penalty0 (9809):\penalty0 2088--2094, 2011.

\bibitem[Dash et~al.(2018)Dash, Ferrari, Malik, Montillo, Maldjian, and
  Wang]{dash2018determining}
Debadatta Dash, Paul Ferrari, Saleem Malik, Albert Montillo, Joseph~A Maldjian,
  and Jun Wang.
\newblock Determining the optimal number of meg trials: A machine learning and
  speech decoding perspective.
\newblock In \emph{Brain Informatics: International Conference, BI 2018,
  Arlington, TX, USA, December 7--9, 2018, Proceedings 11}, pages 163--172.
  Springer, 2018.

\bibitem[Dash et~al.(2019)Dash, Ferrari, Heitzman, and Wang]{dash2019decoding}
Debadatta Dash, Paul Ferrari, Daragh Heitzman, and Jun Wang.
\newblock Decoding speech from single trial meg signals using convolutional
  neural networks and transfer learning.
\newblock In \emph{2019 41st Annual International Conference of the IEEE
  Engineering in Medicine and Biology Society (EMBC)}, pages 5531--5535. IEEE,
  2019.

\bibitem[Dauphin et~al.(2017)Dauphin, Fan, Auli, and Grangier]{glu}
Yann~N. Dauphin, Angela Fan, Michael Auli, and David Grangier.
\newblock Language modeling with gated convolutional networks.
\newblock In \emph{Proceedings of the International Conference on Machine
  Learning}, 2017.

\bibitem[Fernandino et~al.(2022)Fernandino, Tong, Conant, Humphries, and
  Binder]{fernandino2022decoding}
Leonardo Fernandino, Jia-Qing Tong, Lisa~L Conant, Colin~J Humphries, and
  Jeffrey~R Binder.
\newblock Decoding the information structure underlying the neural
  representation of concepts.
\newblock \emph{Proceedings of the National Academy of Sciences}, 119\penalty0
  (6):\penalty0 e2108091119, 2022.

\bibitem[Garc{\'\i}a-Salinas et~al.(2019)Garc{\'\i}a-Salinas,
  Villase{\~n}or-Pineda, Reyes-Garc{\'\i}a, and
  Torres-Garc{\'\i}a]{garcia2019transfer}
Jes{\'u}s~S Garc{\'\i}a-Salinas, Luis Villase{\~n}or-Pineda, Carlos~A
  Reyes-Garc{\'\i}a, and Alejandro~A Torres-Garc{\'\i}a.
\newblock Transfer learning in imagined speech eeg-based bcis.
\newblock \emph{Biomedical Signal Processing and Control}, 50:\penalty0
  151--157, 2019.

\bibitem[Gauthier and Levy(2019)]{gauthier2019linking}
Jon Gauthier and Roger Levy.
\newblock Linking artificial and human neural representations of language.
\newblock \emph{arXiv preprint arXiv:1910.01244}, 2019.

\bibitem[Gramfort et~al.(2013)Gramfort, Luessi, Larson, Engemann, Strohmeier,
  Brodbeck, Goj, Jas, Brooks, Parkkonen, et~al.]{gramfort2013meg}
Alexandre Gramfort, Martin Luessi, Eric Larson, Denis~A Engemann, Daniel
  Strohmeier, Christian Brodbeck, Roman Goj, Mainak Jas, Teon Brooks, Lauri
  Parkkonen, et~al.
\newblock Meg and eeg data analysis with mne-python.
\newblock \emph{Frontiers in neuroscience}, page 267, 2013.

\bibitem[Gwilliams et~al.(2020)Gwilliams, King, Marantz, and
  Poeppel]{gwilliams2022masc}
Laura Gwilliams, Jean-Remi King, Alec Marantz, and David Poeppel.
\newblock Neural dynamics of phoneme sequencing in real speech jointly encode
  order and invariant content.
\newblock \emph{bioRxiv}, 2020.

\bibitem[Gwilliams et~al.(2022)Gwilliams, Flick, Marantz, Pylkkanen, Poeppel,
  and King]{gwilliams2020neural}
Laura Gwilliams, Graham Flick, Alec Marantz, Liina Pylkkanen, David Poeppel,
  and Jean-Remi King.
\newblock Meg-masc: a high-quality magneto-encephalography dataset for
  evaluating natural speech processing.
\newblock \emph{arXiv preprint arXiv:2208.11488}, 2022.

\bibitem[H{\"a}m{\"a}l{\"a}inen et~al.(1993)H{\"a}m{\"a}l{\"a}inen, Hari,
  Ilmoniemi, Knuutila, and Lounasmaa]{hamalainen1993magnetoencephalography}
Matti H{\"a}m{\"a}l{\"a}inen, Riitta Hari, Risto~J Ilmoniemi, Jukka Knuutila,
  and Olli~V Lounasmaa.
\newblock Magnetoencephalography—theory, instrumentation, and applications to
  noninvasive studies of the working human brain.
\newblock \emph{Reviews of modern Physics}, 65\penalty0 (2):\penalty0 413,
  1993.

\bibitem[Haxby et~al.(2001)Haxby, Gobbini, Furey, Ishai, Schouten, and
  Pietrini]{haxby2001distributed}
James~V Haxby, M~Ida Gobbini, Maura~L Furey, Alumit Ishai, Jennifer~L Schouten,
  and Pietro Pietrini.
\newblock Distributed and overlapping representations of faces and objects in
  ventral temporal cortex.
\newblock \emph{Science}, 293\penalty0 (5539):\penalty0 2425--2430, 2001.

\bibitem[Haxby et~al.(2020)Haxby, Guntupalli, Nastase, and
  Feilong]{haxby2020hyperalignment}
James~V Haxby, J~Swaroop Guntupalli, Samuel~A Nastase, and Ma~Feilong.
\newblock Hyperalignment: Modeling shared information encoded in idiosyncratic
  cortical topographies.
\newblock \emph{Elife}, 9:\penalty0 e56601, 2020.

\bibitem[Hendrycks and Gimpel(2016)]{gelu}
Dan Hendrycks and Kevin Gimpel.
\newblock Gaussian error linear units (gelus).
\newblock \emph{arXiv preprint arXiv:1606.08415}, 2016.

\bibitem[Herff et~al.(2015)Herff, Heger, De~Pesters, Telaar, Brunner, Schalk,
  and Schultz]{herff2015brain}
Christian Herff, Dominic Heger, Adriana De~Pesters, Dominic Telaar, Peter
  Brunner, Gerwin Schalk, and Tanja Schultz.
\newblock Brain-to-text: decoding spoken phrases from phone representations in
  the brain.
\newblock \emph{Frontiers in neuroscience}, 9:\penalty0 217, 2015.

\bibitem[Hewitt and Manning(2019)]{hewitt2019structural}
John Hewitt and Christopher~D Manning.
\newblock A structural probe for finding syntax in word representations.
\newblock In \emph{Proceedings of the 2019 Conference of the North American
  Chapter of the Association for Computational Linguistics: Human Language
  Technologies, Volume 1 (Long and Short Papers)}, pages 4129--4138, 2019.

\bibitem[Hickok and Poeppel(2007)]{hickok2007cortical}
Gregory Hickok and David Poeppel.
\newblock The cortical organization of speech processing.
\newblock \emph{Nature reviews neuroscience}, 8\penalty0 (5):\penalty0
  393--402, 2007.

\bibitem[Horikawa and Kamitani(2017)]{horikawa2017generic}
Tomoyasu Horikawa and Yukiyasu Kamitani.
\newblock Generic decoding of seen and imagined objects using hierarchical
  visual features.
\newblock \emph{Nature communications}, 8\penalty0 (1):\penalty0 15037, 2017.

\bibitem[Huth et~al.(2016)Huth, De~Heer, Griffiths, Theunissen, and
  Gallant]{huth2016natural}
Alexander~G Huth, Wendy~A De~Heer, Thomas~L Griffiths, Fr{\'e}d{\'e}ric~E
  Theunissen, and Jack~L Gallant.
\newblock Natural speech reveals the semantic maps that tile human cerebral
  cortex.
\newblock \emph{Nature}, 532\penalty0 (7600):\penalty0 453--458, 2016.

\bibitem[Ide et~al.(2010)Ide, Baker, Fellbaum, and Passonneau]{ide2010manually}
Nancy Ide, Collin~F Baker, Christiane Fellbaum, and Rebecca~J Passonneau.
\newblock The manually annotated sub-corpus: A community resource for and by
  the people.
\newblock In \emph{Proceedings of the ACL 2010 conference short papers}, pages
  68--73, 2010.

\bibitem[Ioffe and Szegedy(2015)]{batchnorm}
Sergey Ioffe and Christian Szegedy.
\newblock Batch normalization: Accelerating deep network training by reducing
  internal covariate shift.
\newblock Technical Report 1502.03167, arXiv, 2015.

\bibitem[Jahangiri et~al.(2018)Jahangiri, Chau, Achanccaray, and
  Sepulveda]{jahangiri2018covert}
Amir Jahangiri, Juan~M Chau, David~R Achanccaray, and Francisco Sepulveda.
\newblock Covert speech vs. motor imagery: a comparative study of class
  separability in identical environments.
\newblock In \emph{2018 40th Annual International Conference of the IEEE
  Engineering in Medicine and Biology Society (EMBC)}, pages 2020--2023. IEEE,
  2018.

\bibitem[Jas et~al.(2017)Jas, Engemann, Bekhti, Raimondo, and
  Gramfort]{jas2017autoreject}
Mainak Jas, Denis~A Engemann, Yousra Bekhti, Federico Raimondo, and Alexandre
  Gramfort.
\newblock Autoreject: Automated artifact rejection for meg and eeg data.
\newblock \emph{NeuroImage}, 159:\penalty0 417--429, 2017.

\bibitem[Jayaram and Barachant(2018)]{jayaram2018moabb}
Vinay Jayaram and Alexandre Barachant.
\newblock Moabb: trustworthy algorithm benchmarking for bcis.
\newblock \emph{Journal of neural engineering}, 15\penalty0 (6):\penalty0
  066011, 2018.

\bibitem[Kamitani and Tong(2005)]{kamitani2005decoding}
Yukiyasu Kamitani and Frank Tong.
\newblock Decoding the visual and subjective contents of the human brain.
\newblock \emph{Nature neuroscience}, 8\penalty0 (5):\penalty0 679--685, 2005.

\bibitem[Kennedy et~al.(2022)Kennedy, Ganesh, and Cervantes]{kennedy2022slow}
Philip Kennedy, A~Ganesh, and AJ~Cervantes.
\newblock Slow firing single units are essential for optimal decoding of silent
  speech.
\newblock 2022.

\bibitem[King et~al.(2013)King, Faugeras, Gramfort, Schurger, El~Karoui, Sitt,
  Rohaut, Wacongne, Labyt, Bekinschtein, et~al.]{king2013single}
Jean-R{\'e}mi King, Fr{\'e}d{\'e}ric Faugeras, Alexandre Gramfort, Aaron
  Schurger, Imen El~Karoui, JD~Sitt, Benjamin Rohaut, C~Wacongne, E~Labyt,
  Tristan Bekinschtein, et~al.
\newblock Single-trial decoding of auditory novelty responses facilitates the
  detection of residual consciousness.
\newblock \emph{Neuroimage}, 83:\penalty0 726--738, 2013.

\bibitem[King et~al.(2018)King, Gwilliams, Holdgraf, Sassenhagen, Barachant,
  Engemann, Larson, and Gramfort]{king2018encoding}
Jean-R{\'e}mi King, Laura Gwilliams, Chris Holdgraf, Jona Sassenhagen,
  Alexandre Barachant, Denis Engemann, Eric Larson, and Alexandre Gramfort.
\newblock Encoding and decoding neuronal dynamics: Methodological framework to
  uncover the algorithms of cognition.
\newblock 2018.

\bibitem[Kingma and Ba(2014)]{ADAM}
Diederik Kingma and Jimmy Ba.
\newblock Adam: A method for stochastic optimization.
\newblock \emph{International Conference on Learning Representations}, 12 2014.

\bibitem[Kohler et~al.(2021)Kohler, Ottenhoff, Goulis, Angrick, Colon, Wagner,
  Tousseyn, Kubben, and Herff]{kohler2021synthesizing}
Jonas Kohler, Maarten~C Ottenhoff, Sophocles Goulis, Miguel Angrick, Albert~J
  Colon, Louis Wagner, Simon Tousseyn, Pieter~L Kubben, and Christian Herff.
\newblock Synthesizing speech from intracranial depth electrodes using an
  encoder-decoder framework.
\newblock \emph{arXiv preprint arXiv:2111.01457}, 2021.

\bibitem[Koizumi et~al.(2018)Koizumi, Ueda, and Nakao]{koizumi2018development}
Koji Koizumi, Kazutaka Ueda, and Masayuki Nakao.
\newblock Development of a cognitive brain-machine interface based on a visual
  imagery method.
\newblock In \emph{2018 40th Annual International Conference of the IEEE
  Engineering in Medicine and Biology Society (EMBC)}, pages 1062--1065. IEEE,
  2018.

\bibitem[Komeiji et~al.(2022)Komeiji, Shigemi, Mitsuhashi, Iimura, Suzuki,
  Sugano, Shinoda, and Tanaka]{komeiji2022transformer}
Shuji Komeiji, Kai Shigemi, Takumi Mitsuhashi, Yasushi Iimura, Hiroharu Suzuki,
  Hidenori Sugano, Koichi Shinoda, and Toshihisa Tanaka.
\newblock Transformer-based estimation of spoken sentences using
  electrocorticography.
\newblock In \emph{ICASSP 2022-2022 IEEE International Conference on Acoustics,
  Speech and Signal Processing (ICASSP)}, pages 1311--1315. IEEE, 2022.

\bibitem[Krishna et~al.(2020)Krishna, Tran, Han, Carnahan, and
  Tewfik]{krishna2020speech}
Gautam Krishna, Co~Tran, Yan Han, Mason Carnahan, and Ahmed~H Tewfik.
\newblock Speech synthesis using eeg.
\newblock In \emph{ICASSP 2020-2020 IEEE International Conference on Acoustics,
  Speech and Signal Processing (ICASSP)}, pages 1235--1238. IEEE, 2020.

\bibitem[K{\"u}bler et~al.(2001)K{\"u}bler, Kotchoubey, Kaiser, Wolpaw, and
  Birbaumer]{kubler2001brain}
Andrea K{\"u}bler, Boris Kotchoubey, Jochen Kaiser, Jonathan~R Wolpaw, and
  Niels Birbaumer.
\newblock Brain--computer communication: Unlocking the locked in.
\newblock \emph{Psychological bulletin}, 127\penalty0 (3):\penalty0 358, 2001.

\bibitem[Lawhern et~al.(2018)Lawhern, Solon, Waytowich, Gordon, Hung, and
  Lance]{lawhern2018eegnet}
Vernon~J Lawhern, Amelia~J Solon, Nicholas~R Waytowich, Stephen~M Gordon,
  Chou~P Hung, and Brent~J Lance.
\newblock Eegnet: a compact convolutional neural network for eeg-based
  brain--computer interfaces.
\newblock \emph{Journal of neural engineering}, 15\penalty0 (5):\penalty0
  056013, 2018.

\bibitem[Lopopolo and van~den Bosch(2020)]{lopopolo2020part}
Alessandro Lopopolo and Antal van~den Bosch.
\newblock Part-of-speech classification from magnetoencephalography data using
  1-dimensional convolutional neural network.
\newblock 2020.

\bibitem[Martin et~al.(2016)Martin, Brunner, Iturrate, Mill{\'a}n, Schalk,
  Knight, and Pasley]{martin2016word}
Stephanie Martin, Peter Brunner, I{\~n}aki Iturrate, Jos{\'e} del~R Mill{\'a}n,
  Gerwin Schalk, Robert~T Knight, and Brian~N Pasley.
\newblock Word pair classification during imagined speech using direct brain
  recordings.
\newblock \emph{Scientific reports}, 6\penalty0 (1):\penalty0 1--12, 2016.

\bibitem[Mermelstein(1976)]{mermelstein_mel}
Paul Mermelstein.
\newblock Distance measures for speech recognition, psychological and
  instrumental.
\newblock \emph{Pattern recognition and artificial intelligence}, 116:\penalty0
  374--388, 1976.

\bibitem[Metzger et~al.(2022)Metzger, Liu, Moses, Dougherty, Seaton,
  Littlejohn, Chartier, Anumanchipalli, Tu-Chan, Ganguly,
  et~al.]{metzger2022generalizable}
Sean~L Metzger, Jessie~R Liu, David~A Moses, Maximilian~E Dougherty, Margaret~P
  Seaton, Kaylo~T Littlejohn, Josh Chartier, Gopala~K Anumanchipalli, Adelyn
  Tu-Chan, Karunesh Ganguly, et~al.
\newblock Generalizable spelling using a speech neuroprosthesis in an
  individual with severe limb and vocal paralysis.
\newblock \emph{Nature Communications}, 13\penalty0 (1):\penalty0 6510, 2022.

\bibitem[Millet and Dunbar(2022)]{millet2022self}
Juliette Millet and Ewan Dunbar.
\newblock Do self-supervised speech models develop human-like perception
  biases?
\newblock \emph{arXiv preprint arXiv:2205.15819}, 2022.

\bibitem[Millet et~al.(2022)Millet, Caucheteux, Orhan, Boubenec, Gramfort,
  Dunbar, Pallier, and King]{millet2022toward}
Juliette Millet, Charlotte Caucheteux, Pierre Orhan, Yves Boubenec, Alexandre
  Gramfort, Ewan Dunbar, Christophe Pallier, and Jean-Remi King.
\newblock Toward a realistic model of speech processing in the brain with
  self-supervised learning.
\newblock \emph{arXiv preprint arXiv:2206.01685}, 2022.

\bibitem[Miyawaki et~al.(2008)Miyawaki, Uchida, Yamashita, Sato, Morito,
  Tanabe, Sadato, and Kamitani]{miyawaki2008visual}
Yoichi Miyawaki, Hajime Uchida, Okito Yamashita, Masa-aki Sato, Yusuke Morito,
  Hiroki~C Tanabe, Norihiro Sadato, and Yukiyasu Kamitani.
\newblock Visual image reconstruction from human brain activity using a
  combination of multiscale local image decoders.
\newblock \emph{Neuron}, 60\penalty0 (5):\penalty0 915--929, 2008.

\bibitem[Moinnereau et~al.(2018)Moinnereau, Brienne, Brodeur, Rouat,
  Whittingstall, and Plourde]{moinnereau2018classification}
Marc-Antoine Moinnereau, Thomas Brienne, Simon Brodeur, Jean Rouat, Kevin
  Whittingstall, and Eric Plourde.
\newblock Classification of auditory stimuli from eeg signals with a regulated
  recurrent neural network reservoir.
\newblock \emph{arXiv preprint arXiv:1804.10322}, 2018.

\bibitem[Moses et~al.(2021)Moses, Metzger, Liu, Anumanchipalli, Makin, Sun,
  Chartier, Dougherty, Liu, Abrams, et~al.]{moses2021neuroprosthesis}
David~A Moses, Sean~L Metzger, Jessie~R Liu, Gopala~K Anumanchipalli, Joseph~G
  Makin, Pengfei~F Sun, Josh Chartier, Maximilian~E Dougherty, Patricia~M Liu,
  Gary~M Abrams, et~al.
\newblock Neuroprosthesis for decoding speech in a paralyzed person with
  anarthria.
\newblock \emph{New England Journal of Medicine}, 385\penalty0 (3):\penalty0
  217--227, 2021.

\bibitem[Murphy et~al.(2022)Murphy, Bohnet, McDonald, and
  Noppeney]{murphy2022decoding}
Alex Murphy, Bernd Bohnet, Ryan McDonald, and Uta Noppeney.
\newblock Decoding part-of-speech from human eeg signals.
\newblock In \emph{Proceedings of the 60th Annual Meeting of the Association
  for Computational Linguistics (Volume 1: Long Papers)}, pages 2201--2210,
  2022.

\bibitem[Nguyen et~al.(2017)Nguyen, Karavas, and
  Artemiadis]{nguyen2017inferring}
Chuong~H Nguyen, George~K Karavas, and Panagiotis Artemiadis.
\newblock Inferring imagined speech using eeg signals: a new approach using
  riemannian manifold features.
\newblock \emph{Journal of neural engineering}, 15\penalty0 (1):\penalty0
  016002, 2017.

\bibitem[Nishimoto et~al.(2011)Nishimoto, Vu, Naselaris, Benjamini, Yu, and
  Gallant]{nishimoto2011reconstructing}
Shinji Nishimoto, An~T Vu, Thomas Naselaris, Yuval Benjamini, Bin Yu, and
  Jack~L Gallant.
\newblock Reconstructing visual experiences from brain activity evoked by
  natural movies.
\newblock \emph{Current biology}, 21\penalty0 (19):\penalty0 1641--1646, 2011.

\bibitem[Orpella et~al.(2022)Orpella, Mantegna, Assaneo, and
  Poeppel]{orpella2022speech}
Joan Orpella, Francesco Mantegna, Florencia Assaneo, and David Poeppel.
\newblock Speech imagery decoding as a window to speech planning and
  production.
\newblock \emph{bioRxiv}, pages 2022--05, 2022.

\bibitem[Ott et~al.(2019)Ott, Edunov, Baevski, Fan, Gross, Ng, Grangier, and
  Auli]{ott2019fairseq}
Myle Ott, Sergey Edunov, Alexei Baevski, Angela Fan, Sam Gross, Nathan Ng,
  David Grangier, and Michael Auli.
\newblock fairseq: A fast, extensible toolkit for sequence modeling.
\newblock In \emph{Proceedings of NAACL-HLT 2019: Demonstrations}, 2019.
\newblock URL
  \url{https://github.com/pytorch/fairseq/blob/main/examples/wav2vec}.

\bibitem[Owen et~al.(2006)Owen, Coleman, Boly, Davis, Laureys, and
  Pickard]{owen2006detecting}
Adrian~M Owen, Martin~R Coleman, Melanie Boly, Matthew~H Davis, Steven Laureys,
  and John~D Pickard.
\newblock Detecting awareness in the vegetative state.
\newblock \emph{science}, 313\penalty0 (5792):\penalty0 1402--1402, 2006.

\bibitem[Ozcelik and VanRullen(2023)]{ozcelik2023brain}
Furkan Ozcelik and Rufin VanRullen.
\newblock Brain-diffuser: Natural scene reconstruction from fmri signals using
  generative latent diffusion.
\newblock \emph{arXiv preprint arXiv:2303.05334}, 2023.

\bibitem[Panachakel and Ramakrishnan(2021)]{panachakel2021decoding}
Jerrin~Thomas Panachakel and Angarai~Ganesan Ramakrishnan.
\newblock Decoding covert speech from eeg-a comprehensive review.
\newblock \emph{Frontiers in Neuroscience}, 15:\penalty0 392, 2021.

\bibitem[Pascual et~al.(2022)Pascual, Egressy, Affolter, Cai, Richter, and
  Wattenhofer]{pascual2022improving}
Damian Pascual, B{\'e}ni Egressy, Nicolas Affolter, Yiming Cai, Oliver Richter,
  and Roger Wattenhofer.
\newblock Improving brain decoding methods and evaluation.
\newblock In \emph{ICASSP 2022-2022 IEEE International Conference on Acoustics,
  Speech and Signal Processing (ICASSP)}, pages 1476--1480. IEEE, 2022.

\bibitem[Pasley et~al.(2012)Pasley, David, Mesgarani, Flinker, Shamma, Crone,
  Knight, and Chang]{pasley2012reconstructing}
Brian~N Pasley, Stephen~V David, Nima Mesgarani, Adeen Flinker, Shihab~A
  Shamma, Nathan~E Crone, Robert~T Knight, and Edward~F Chang.
\newblock Reconstructing speech from human auditory cortex.
\newblock \emph{PLoS biology}, 10\penalty0 (1):\penalty0 e1001251, 2012.

\bibitem[Pedregosa et~al.(2011)Pedregosa, Varoquaux, Gramfort, Michel, Thirion,
  Grisel, Blondel, Prettenhofer, Weiss, Dubourg, et~al.]{pedregosa2011scikit}
Fabian Pedregosa, Ga{\"e}l Varoquaux, Alexandre Gramfort, Vincent Michel,
  Bertrand Thirion, Olivier Grisel, Mathieu Blondel, Peter Prettenhofer, Ron
  Weiss, Vincent Dubourg, et~al.
\newblock Scikit-learn: Machine learning in python.
\newblock \emph{the Journal of machine Learning research}, 12:\penalty0
  2825--2830, 2011.
\newblock URL \url{https://scikit-learn.org/}.

\bibitem[Pei et~al.(2011)Pei, Barbour, Leuthardt, and Schalk]{pei2011decoding}
Xiaomei Pei, Dennis~L Barbour, Eric~C Leuthardt, and Gerwin Schalk.
\newblock Decoding vowels and consonants in spoken and imagined words using
  electrocorticographic signals in humans.
\newblock \emph{Journal of neural engineering}, 8\penalty0 (4):\penalty0
  046028, 2011.

\bibitem[Pels et~al.(2017)Pels, Aarnoutse, Ramsey, and
  Vansteensel]{pels2017estimated}
Elmar~GM Pels, Erik~J Aarnoutse, Nick~F Ramsey, and Mariska~J Vansteensel.
\newblock Estimated prevalence of the target population for brain-computer
  interface neurotechnology in the netherlands.
\newblock \emph{Neurorehabilitation and neural repair}, 31\penalty0
  (7):\penalty0 677--685, 2017.

\bibitem[Radford et~al.(2021)Radford, Kim, Hallacy, Ramesh, Goh, Agarwal,
  Sastry, Askell, Mishkin, Clark, et~al.]{radford2021learning}
Alec Radford, Jong~Wook Kim, Chris Hallacy, Aditya Ramesh, Gabriel Goh,
  Sandhini Agarwal, Girish Sastry, Amanda Askell, Pamela Mishkin, Jack Clark,
  et~al.
\newblock Learning transferable visual models from natural language
  supervision.
\newblock In \emph{International Conference on Machine Learning}, pages
  8748--8763. PMLR, 2021.

\bibitem[Roy et~al.(2019)Roy, Banville, Albuquerque, Gramfort, Falk, and
  Faubert]{roy2019deep}
Yannick Roy, Hubert Banville, Isabela Albuquerque, Alexandre Gramfort, Tiago~H
  Falk, and Jocelyn Faubert.
\newblock Deep learning-based electroencephalography analysis: a systematic
  review.
\newblock \emph{Journal of neural engineering}, 16\penalty0 (5):\penalty0
  051001, 2019.

\bibitem[Schirrmeister et~al.(2017)Schirrmeister, Springenberg, Fiederer,
  Glasstetter, Eggensperger, Tangermann, Hutter, Burgard, and
  Ball]{schirrmeister2017deep}
Robin~Tibor Schirrmeister, Jost~Tobias Springenberg, Lukas Dominique~Josef
  Fiederer, Martin Glasstetter, Katharina Eggensperger, Michael Tangermann,
  Frank Hutter, Wolfram Burgard, and Tonio Ball.
\newblock Deep learning with convolutional neural networks for eeg decoding and
  visualization.
\newblock \emph{Human brain mapping}, 38\penalty0 (11):\penalty0 5391--5420,
  2017.

\bibitem[Schoffelen et~al.(2019)Schoffelen, Oostenveld, Lam, Udd{\'e}n,
  Hult{\'e}n, and Hagoort]{Schoffelen_19_204subjectMultimodalNeuroimaging}
Jan-Mathijs Schoffelen, Robert Oostenveld, Nietzsche H.~L. Lam, Julia
  Udd{\'e}n, Annika Hult{\'e}n, and Peter Hagoort.
\newblock A 204-subject multimodal neuroimaging dataset to study language
  processing.
\newblock \emph{Scientific Data}, 6\penalty0 (1):\penalty0 17, April 2019.
\newblock ISSN 2052-4463.
\newblock \doi{10.1038/s41597-019-0020-y}.
\newblock URL
  \url{https://data.donders.ru.nl/collections/di/dccn/DSC_3011020.09_236?0}.

\bibitem[Schwenk and Douze(2017)]{schwenk2017learning}
Holger Schwenk and Matthijs Douze.
\newblock Learning joint multilingual sentence representations with neural
  machine translation.
\newblock \emph{arXiv preprint arXiv:1704.04154}, 2017.
\newblock URL \url{https://github.com/facebookresearch/laser}.

\bibitem[Speer(2022)]{robyn_speer_2022_7199437}
Robyn Speer.
\newblock rspeer/wordfreq: v3.0.
\newblock September 2022.
\newblock \doi{10.5281/zenodo.7199437}.
\newblock URL \url{https://doi.org/10.5281/zenodo.7199437}.

\bibitem[Sree and Kavitha(2017)]{sree2017vowel}
R~Anandha Sree and A~Kavitha.
\newblock Vowel classification from imagined speech using sub-band eeg
  frequencies and deep belief networks.
\newblock In \emph{2017 fourth international conference on signal processing,
  communication and networking (ICSCN)}, pages 1--4. IEEE, 2017.

\bibitem[Stanger and Cawley(1996)]{stanger1996demographics}
Carol~A Stanger and Michael~F Cawley.
\newblock Demographics of rehabilitation robotics users.
\newblock \emph{Technology and Disability}, 5\penalty0 (2):\penalty0 125--137,
  1996.

\bibitem[Stavisky et~al.(2018)Stavisky, Rezaii, Willett, Hochberg, Shenoy, and
  Henderson]{stavisky2018decoding}
Sergey~D Stavisky, Paymon Rezaii, Francis~R Willett, Leigh~R Hochberg,
  Krishna~V Shenoy, and Jaimie~M Henderson.
\newblock Decoding speech from intracortical multielectrode arrays in dorsal
  “arm/hand areas” of human motor cortex.
\newblock In \emph{2018 40th Annual International Conference of the IEEE
  Engineering in Medicine and Biology Society (EMBC)}, pages 93--97. IEEE,
  2018.

\bibitem[Sun and Qin(2016)]{sun2016neural}
Pengfei Sun and Jun Qin.
\newblock Neural networks based eeg-speech models.
\newblock \emph{arXiv preprint arXiv:1612.05369}, 2016.

\bibitem[Tang et~al.(2022)Tang, LeBel, Jain, and Huth]{tang2022semantic}
Jerry Tang, Amanda LeBel, Shailee Jain, and Alexander~G Huth.
\newblock Semantic reconstruction of continuous language from non-invasive
  brain recordings.
\newblock \emph{bioRxiv}, pages 2022--09, 2022.

\bibitem[Thomas et~al.(2022)Thomas, R{\'e}, and Poldrack]{thomas2022self}
Armin~W Thomas, Christopher R{\'e}, and Russell~A Poldrack.
\newblock Self-supervised learning of brain dynamics from broad neuroimaging
  data.
\newblock \emph{arXiv preprint arXiv:2206.11417}, 2022.

\bibitem[Vaidya et~al.(2022)Vaidya, Jain, and Huth]{vaidya2022self}
Aditya~R Vaidya, Shailee Jain, and Alexander~G Huth.
\newblock Self-supervised models of audio effectively explain human cortical
  responses to speech.
\newblock \emph{arXiv preprint arXiv:2205.14252}, 2022.

\bibitem[Willett et~al.(2021)Willett, Avansino, Hochberg, Henderson, and
  Shenoy]{willett2021high}
Francis~R Willett, Donald~T Avansino, Leigh~R Hochberg, Jaimie~M Henderson, and
  Krishna~V Shenoy.
\newblock High-performance brain-to-text communication via handwriting.
\newblock \emph{Nature}, 593\penalty0 (7858):\penalty0 249--254, 2021.

\bibitem[Xu et~al.(2012)Xu, Lorbert, Ramadge, Guntupalli, and
  Haxby]{xu2012regularized}
Hao Xu, Alexander Lorbert, Peter~J Ramadge, J~Swaroop Guntupalli, and James~V
  Haxby.
\newblock Regularized hyperalignment of multi-set fmri data.
\newblock In \emph{2012 IEEE Statistical Signal Processing Workshop (SSP)},
  pages 229--232. IEEE, 2012.

\bibitem[Yang et~al.(2021)Yang, Hira, Ni, Chourdia, Astafurov, Chen, Yeh,
  Puhrsch, Pollack, Genzel, Greenberg, Yang, Lian, Mahadeokar, Hwang, Chen,
  Goldsborough, Roy, Narenthiran, Watanabe, Chintala, Quenneville-Bélair, and
  Shi]{yang2021torchaudio}
Yao-Yuan Yang, Moto Hira, Zhaoheng Ni, Anjali Chourdia, Artyom Astafurov,
  Caroline Chen, Ching-Feng Yeh, Christian Puhrsch, David Pollack, Dmitriy
  Genzel, Donny Greenberg, Edward~Z. Yang, Jason Lian, Jay Mahadeokar, Jeff
  Hwang, Ji~Chen, Peter Goldsborough, Prabhat Roy, Sean Narenthiran, Shinji
  Watanabe, Soumith Chintala, Vincent Quenneville-Bélair, and Yangyang Shi.
\newblock Torchaudio: Building blocks for audio and speech processing.
\newblock \emph{arXiv preprint arXiv:2110.15018}, 2021.

\bibitem[Young et~al.(2002)Young, Evermann, Gales, Hain, Kershaw, Liu, Moore,
  Odell, Ollason, Povey, et~al.]{young2002htk}
Steve Young, Gunnar Evermann, Mark Gales, Thomas Hain, Dan Kershaw, Xunying
  Liu, Gareth Moore, Julian Odell, Dave Ollason, Dan Povey, et~al.
\newblock The htk book.
\newblock \emph{Cambridge university engineering department}, 3\penalty0
  (175):\penalty0 12, 2002.

\end{thebibliography}
\bibliographystyle{plainnat}

\clearpage
\appendix
\numberwithin{equation}{section}
\counterwithin{table}{section}
\counterwithin{figure}{section}

\section{Supplementary information}

\subsection{Datasets}
\label{supp:datasets}

The data from \citet{Schoffelen_19_204subjectMultimodalNeuroimaging}
 was provided (in part) by the Donders Institute for Brain, Cognition and Behaviour with a ``RU-DI-HD-1.0'' licence.%
 The data for \citet{gwilliams2020neural} is available under
 CC0 1.0 Universal. %
 The data for \citet{broderick2018electrophysiological}
 is available under the same licence. %
 Finally, the data from \citet{brennan2019hierarchical}
 is available under the CC BY 4.0 licence. %
All audio files were provided by the authors of each dataset.

\begin{table}[H]
\red
\centering
\caption{\textbf{Top-1 Accuracy.} Segment-level accuracy related to Table \ref{tab:results}.}

\begin{tabular}{lllll}
\toprule
Model &    Brennan (EEG) &  Broderick (EEG) &   Gwilliams (MEG) &  Schoffelen (MEG) \\
\midrule
Random model  &         0.5$\pm$0.0 &         0.1$\pm$0.0 &          0.1$\pm$0.0 &          0.1$\pm$0.0 \\
Base Model    &         0.7$\pm$0.2 &         0.1$\pm$0.0 &          3.0$\pm$0.3 &          5.8$\pm$0.6 \\
+ CLIP        &         0.9$\pm$0.6 &         2.1$\pm$0.4 &         26.2$\pm$0.6 &         25.1$\pm$0.6 \\
+ Deep Mel    &  \textbf{5.2}$\pm$1.1 &         4.2$\pm$0.7 &         34.8$\pm$1.2 &         31.6$\pm$1.0 \\
+ wav2vec 2.0 &         \textbf{5.2}$\pm$0.8 &  \textbf{5.0}$\pm$0.4 &  \textbf{41.3}$\pm$0.1 &  \textbf{36.8}$\pm$0.4 \\
\bottomrule
\end{tabular}\label{tab:results_top1}
\color{black}
\end{table}

\subsection{Impact of clamping}\label{sec:clamping}

\jr{Clamping is here motivated by the fact that electro-magnetic recordings can be subject to perturbations orders of magnitudes larger than brain activity. As explained in Section~\ref{sec:experiments}, clamping is performed as follows: for each recording independently, we apply a robust scaler such that the data range [-1, 1] maps to the [0.25, 0.75] quantile range. The resulting M/EEG signals are thus expected to have a scale of the order of 1. In Table~\ref{tab:clamp}, we provide the top-10 accuracy for our model similar to Table~\ref{tab:results}. Extending the clamping range to 100 standard deviations does not appear to help extracting more information. This result suggests that clamping effectively takes care of outlier without removing useful information. On the contrary, the removal of clamping leads to a collapse of decoding performance. This is expected, as extreme outliers will impact, for instance, the BatchNorm mean and standard deviation estimates, and one outlier can thus impact the entire batch. Outliers can also cause extreme gradients and throw off the optimization process. Overall, these analyses emphasize the importance of using clamping for M/EEG analyses. }{}

\begin{table}[H]
\red
\caption{\textbf{Clamping results.} Top-10 segment-level accuracy related to Table \ref{tab:results}.}
\begin{center}
    \resizebox{0.98\textwidth}{!}{

\begin{tabular}{lcccc}
\toprule
Clamping &     Brennan (EEG) &   Broderick (EEG) &   Gwilliams (MEG) &  Schoffelen (MEG) \\
\midrule
20    &         \textbf{25.7} $\pm$ 2.9 &  \textbf{17.7} $\pm$ 0.6 &  \textbf{70.7} $\pm$ 0.1 &  \textbf{67.5} $\pm$ 0.4 \\
100   &  \textbf{27.1} $\pm$ 2.6 &         \textbf{17.6} $\pm$ 0.0 &         \textbf{70.6} $\pm$ 0.3 &        \textbf{67.2} $\pm$ 0.9 \\
None  &         14.1 $\pm$ 1.0 &          0.5 $\pm$ 0.0 &        23.6 $\pm$ 24.6 &          1.5 $\pm$ 0.3 \\
\bottomrule
\end{tabular}
}
\end{center}
\label{tab:clamp}
\color{black}
\end{table}

\begin{figure}[ht]
  \centering
  \includegraphics[width=\textwidth]{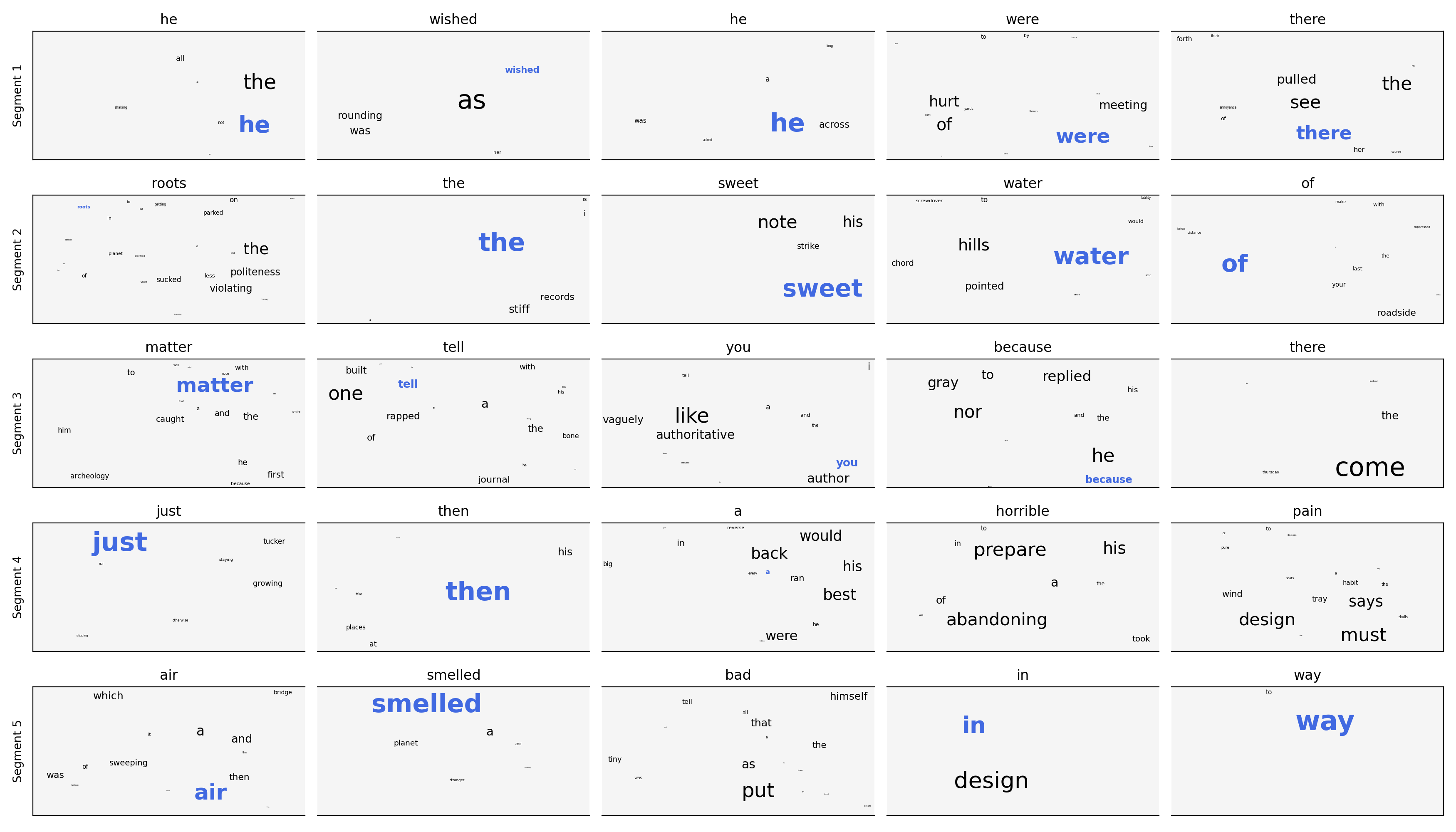}
  \caption{
    \jr{Similar illustration to Figure \ref{fig:results_thankyou} but for five representative speech segments. The top and bottom segments are the easiest and hardest to decode, respectively. For each segment, we plot the predictions obtained for the subject with the median decoding scores across the cohort.}{}
}

  \label{fig:results_sentences}
\end{figure}

\subsection{Comparison with \texttt{Autoreject}}\label{sec:autoreject}

\jr{To evaluate the potential usefulness of advanced M/EEG preprocessing techniques, we compare our model to one trained on M/EEG data preprocess with the \texttt{Autoreject} package~\citep{jas2017autoreject}. This package aims to detect and correct corrupted channels based on their spatial neighborhood. Note that this package can also reject samples that are too corrupted. However, as this procedure would change the definition of the test set, we do not consider it here. Due to the added complexity of applying Autoreject, which requires in particular the loading of the full dataset in memory, we only evaluated it on the first 16 recordings of \citet{gwilliams2020neural} Overall, similar performances were obtained with and without Autoreject. This result suggests that our model does not trivially benefit from advanced preprocessing techniques.}{}

\begin{center}
\red
\begin{tabular}{lr}
\toprule
        \textbf{Method}     & \textit{Gwilliams2022} (16 rec.) \\ 
 \midrule
 clamping at 20 & 54.7 $\pm$ 1.72\\
 autoreject \citep{jas2017autoreject} & 53.4 $\pm$ 1.51\\
\bottomrule
\end{tabular}
\color{black}
\end{center}

\subsection{Impact of EEG/MEG time offset}

\jr{Our model uses a fixed delay of 150\,ms to align speech and EEG/MEG representations.
This decision was originally motivated by the non-compressible delay that exists between the cochlea's response and the cortical activations. To further investigate this decision choice, we here study the impact of this delay on the Gwilliams2022 dataset. We observe a small impact of this parameter, although setting it to 0 only reduces the top-10 accuracy by 0.5\%.}{}

\begin{center}
\red
\begin{tabular}{lr}
\toprule
        \textbf{Delay (ms)}     & \textit{Gwilliams2022} \\ 
 \midrule
0          &  69.9 $\pm$ 0.27\\
50         &  70.1 $\pm$ 0.13\\
100        &  \textbf{70.4} $\pm$ 0.05 \\
150        &  \textbf{70.4} $\pm$ 0.11\\
200        &  69.6 $\pm$ 0.59\\
250        &  68.5 $\pm$ 0.14\\
300        &  67.6 $\pm$ 0.17 \\
\bottomrule
\end{tabular}
\color{black}
\end{center}

\subsection{Impact of the number of Mels}

\jr{Are the models trained to decode Mel features (as opposed to latent representations) impeded by the number of Mel?}{} To study this issue, we evaluate the impact of \jr{the number of frequency bands used for the Mel spectrogram}{} for different versions of the model\jr{, while keeping the minimum and maximum frequencies fixed}{}. For clarity, we only provide the average top-10 accuracy overall datasets. We observe a small increase of the accuracy when using more Mel bands \jr{for all the models}{}.

\begin{center}
\red
\begin{tabular}{lrrrr}
\toprule
 & \multicolumn{4}{c}{\textbf{\# Mel bands}} \\
        \textbf{Model}     & 20 & 40 & 80 & 120 \\
 \midrule
 Base model & 9.3 &   9.8 &   9.8 &  10.0 \\
 + CLIP & 30.5 &  30.8 &  31.2 &  31.8 \\
 + Deep Mel & 40.0 &  40.5 &  38.2 &  41.3 \\
\bottomrule
\end{tabular}
\color{black}
\end{center}

\subsection{Impact of the batch size and learning rate}

\jr{We evaluate the impact of batch size and learning rate on the model performance. For simplicity, these evaluations are reported for the \citet{gwilliams2020neural} dataset only. For the main study, we used a batch size of 256 and a learning rate of $3\mathrm{e}{-}4$. We run an evaluation of a grid of 4 learning rates and 4 batch sizes. Overall, we observe better results with larger batch sizes, and we observe instabilities with larger learning rates. }{}%

\begin{center}
\red
\begin{tabular}{lrrrr}
\toprule
 & \multicolumn{4}{c}{\textbf{Batch size}} \\
        \textbf{Learning rate}     & 32 & 64 & 128 & 256 \\
 \midrule
 $1\mathrm{e}{-}4$ &  65.1 $\pm$ 0.62 &  68.1 $\pm$ 0.29 &  68.2 $\pm$ 0.19 &   68.5 $\pm$ 0.28\\
 $3\mathrm{e}{-}4$ & 62.5 $\pm$ 0.59 &  65.8 $\pm$ 0.45 &  68.5 $\pm$ 0.59 &   70.4 $\pm$ 0.11 \\
 $6\mathrm{e}{-}4$ & 20.5 $\pm$ 34.2 &   0.7 $\pm$ 0.02 &  67.0 $\pm$ 0.25 &   68.7 $\pm$ 0.22 \\
 $1\mathrm{e}{-}3$ & 0.7 $\pm$ 0.0 &   0.7 $\pm$ 0.02 &  22.3 $\pm$ 37.3 &  36.9 $\pm$ 32.0 \\
\bottomrule
\end{tabular}
\color{black}
\end{center}

\subsection{Zero-shot decoding of words}

\jrr{The contrastive objective allows our model to select the most likely speech sounds of a test set given brain activity. To what extent does this approach allow the decoding of words absent from the training set? To evaluate this issue, we report the top-10 accuracy at the word level, depending on whether the words are (1) present in or (2) absent from the training set, respectively. Overall, the results show that for EEG, such zero-shot decoding of words falls dramatically. For MEG, however, such zero-shot decoding of words is remarkably close to the accuracy obtained for words present in the training set. These elements strengthen the superiority of MEG over EEG, and show that our approach can be efficiently used to decode words never present during training.}{}

\begin{center}
\redd

\begin{tabular}{lrrrr}
\toprule
 &  Brennan &  Broderick &  Gwilliams &  Schoffelen \\
 & (EEG) &  (EEG) &  (MEG) &  (MEG) \\
\midrule
Words in test  &            9.7 &              8.0 &             64.0 &              61.6 \\
Words in train &           35.9 &             32.9 &             70.0 &              73.4 \\
\bottomrule
\end{tabular}
\color{black}
\end{center}

\subsection{Decoding of isolated words}

\jr{To what extent can our approach be used to decode words presented in isolation? To explore this issue, we evaluated our model using a subset from \citet{Schoffelen_19_204subjectMultimodalNeuroimaging}, where subjects are presented with random word lists. We use a segment ranging from -300\,ms to +500\,ms relative to word onset. The results, displayed in Supplementary Figure \ref{fig:single}, show that 
our model reaches a zero-shot top-1 accuracy of 22.7\% when we limit the vocabulary size to 50 at test set, with subjects peaking at 42.9\%. While this performance is low, it is interesting to compare it to \citet{moses2021neuroprosthesis} who report a top-1 accuracy of 39.5\% with a model trained to decode the production of individual words, without the use of a language model, \emph{i.e.}  independently of context.}{}

\begin{figure}[ht]
  \centering
  \includegraphics[width=\textwidth]{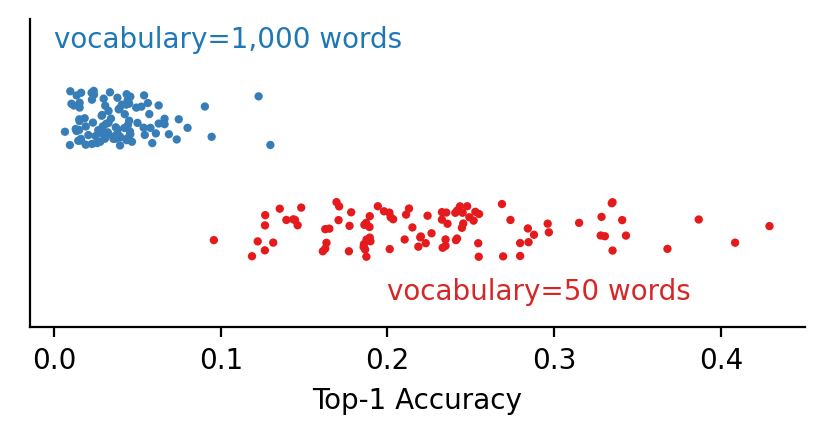}
  \caption{
    \jr{Decoding performance for single-word, using a vocabulary of 1,000 (red) or 50 (blue) words.}{}
}\label{fig:single}
\end{figure}

\red
\subsection{Spatial attention}

\begin{figure}[ht]
  \centering
  \includegraphics[width=\textwidth]{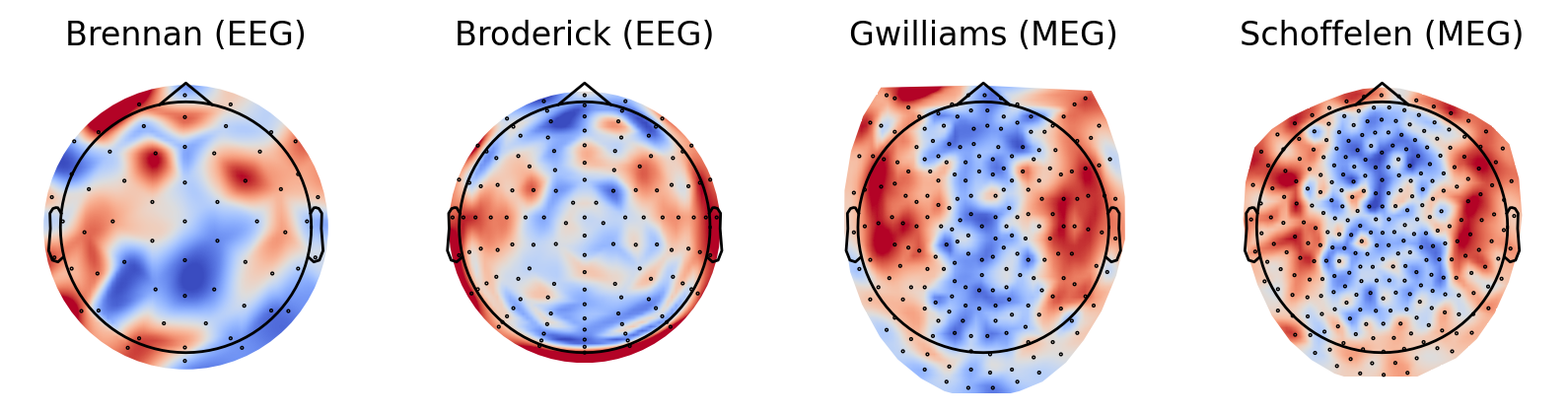}
  \caption{\jr{\textbf{Attention weights.} Red color indicate that the M/EEG sensors is, on average, associated with a higher spatial attention weight. At the exception of the Brennan dataset, the topographies highlight channels typically activated during auditory stimulation.}{}
}

  \label{fig:attention}
\end{figure}

We provide a visualisation of the spatial attention described in Section~\ref{sec:brain_encoder} in Figure~\ref{fig:attention}.

\color{black}

\end{document}